\documentclass[superscriptaddress, prd, aps,amsmath,amssymb,showpacs,showkeys, twocolumn,geometry]{revtex4-1}
\usepackage{graphicx,color}
\usepackage{epstopdf}   % lets pdflatex use the .eps figures directly
\graphicspath{{figures/}}
\usepackage{subfig}
\usepackage{times}
\usepackage{xcolor}
\usepackage{amsmath}
\usepackage{booktabs}
\newcommand{\be}{\begin{equation}}
\newcommand{\ee}{\end{equation}}
\newcommand{\beq}{\begin{eqnarray}}
\newcommand{\eeq}{\end{eqnarray}}
\usepackage[%
  colorlinks=true,
  urlcolor=blue,
  linkcolor=red,
  citecolor=blue
]{hyperref}
\usepackage{orcidlink}

\usepackage{geometry}
\begin{document}

\title{Dirac stars in Einstein-Euler-Heisenberg theory}

\author{Salvatore Capozziello}
\email{capozziello@na.infn.it}
\affiliation{Dipartimento di Fisica ``E. Pancini", Universit\`a di Napoli ``Federico II", Complesso Universitario di Monte Sant’ Angelo, Edificio G, Via Cinthia, I-80126, Napoli, Italy,}
\affiliation{Istituto Nazionale di Fisica Nucleare (INFN), sez. di Napoli, Via Cinthia 9, I-80126 Napoli, Italy,}
\affiliation{Scuola Superiore Meridionale, Via Mezzocannone 4, I-80134, Napoli, Italy.}

\author{Allah Ditta\orcidlink{0000-0001-7758-8736}}
\email{mradshahid01@gmail.com}
\affiliation{Inha University in Tashkent, Ziyolilar 9 Tashkent 100170, Uzbekistan}
\affiliation{Department of Mathematics, School of Science, University of Management and Technology, Lahore, 54000, Pakistan.}

\author{M. Yousaf\orcidlink{0000-0001-7280-177X}}
\email{myousaf.math@gmail.com}
\affiliation{Department of Mathematics, Virtual University of Pakistan, 54-Lawrence Road, Lahore 54000, Pakistan.}

\author{G. Mustafa}
\email{gmustafa3828@gmail.com }
\affiliation{Department of Physics, Zhejiang Normal University,
Jinhua 321004, China}

\begin{abstract}
We study self-gravitating configurations of a charged classical spinor field coupled to gravity and to Euler-Heisenberg (EH) nonlinear electrodynamics, truncated at quartic order. The spinor is gauged under a U(1) field and may carry a quartic self-interaction. For static and spherically symmetric configurations we derive the field equations, reduce them to a dimensionless eigenvalue problem for the spinor frequency and solve it numerically. In the uncharged limit our solutions coincide with the known Dirac stars of general relativity. The electric charge increases the frequency, the mass and the size of the stars, and the maximal mass grows from $0.709$ to $1.027$ (in units of $M_{\rm Pl}^2/\mu$) when the charge coupling rises from zero to $0.75$. The quartic term changes the relation between the electric field and the enclosed charge, so that the field inside the star is weaker than in the Maxwell case. Our central result concerns stars with the same number of fermions, and therefore the same total charge. For them the quartic term changes the mass by less than $0.2\%$, but it makes the stars more compact and more strongly bound and shifts the turning point of the family to a smaller particle number. This shift is what lowers the maximal mass, by $9\%$ (from $0.924$ to $0.840$) at charge coupling $0.6$ as the dimensionless EH coupling goes from zero to $200$. In the weak-field regime of quantum electrodynamics (QED) the relative mass correction is $-1.5\times10^{-5}(E_{\max}/E_c)^2$, so the larger couplings describe the quartic model as an effective nonlinear electrodynamics rather than a controlled QED correction. All solutions satisfy the null, weak, strong and dominant energy conditions, show a positive pressure anisotropy, and remain free of horizons. Their stability is discussed through the binding energy and the turning point of the mass curve.
\end{abstract}

\maketitle

\section{Introduction}

Existence and gravitational dynamics of compact configurations supported by fundamental fields constitute an important direction in relativistic astrophysics as well as theoretical gravitation. Beyond conventional description of compact stars in terms of perfect or anisotropic fluids, self-gravitating scalar, vector, and spinor fields provide a complementary scenario for investigating interplay between matter interactions, spacetime curvature, as well as gravitational equilibrium. Such configurations are particularly interesting because they can form regular, spatially localized, as well as asymptotically flat solutions without necessarily developing an event horizon, while theoretical foundations of this subject established by Kaup \cite{Kaup1968}, Ruffini and his collaborator Bonazzola \cite{Ruffini1969}, who demonstrated existence of self-gravitating scalar field configurations, subsequently known as boson stars. Their equilibrium properties, dynamical behavior, stability, along with potential astrophysical relevance investigated extensively in \cite{Schunck2003,Liebling2023}. Extension of this scenario to massive vector fields led to construction of Proca stars as presented in \cite{Brito2016}, illustrating that gravitationally bound configurations are not restricted to scalar matter. Among these alternatives, self-gravitating spinor configurations occupy a distinctive position because their equilibrium properties are governed by coupled dynamics of fermionic degrees of freedom as well as spacetime geometry, while existence of particle-like solutions to Einstein Dirac equations established by Finster, and his collaborators in \cite{Finster1999a}. In their construction, two spinor states with opposite angular momenta employed to obtain a spherically symmetric energy momentum tensor compatible with a static gravitational background, subsequent investigations developed a unified description of scalar, Dirac, and Proca stars, highlighting similarities as well as differences between their families of equilibrium solutions as presented in \cite{Herdeiro2017}. Rotating generalizations further demonstrated richness of self-gravitating spinor systems as well as their possible spacetime geometries \cite{Herdeiro2019}. 
Further investigations of self-interacting scalar, Dirac and Proca stars highlighted role of nonlinear matter
interactions in determining regular equilibrium configurations \cite{Herdeiro2020SI}, while semiclassical formulations of Einstein-Dirac system also provided an alternative perspective on self-gravitating fermionic matter in \cite{Kain2023}. Moreover, rapidly rotating Dirac stars \cite{Dzhunushaliev2022Rot}, Dirac-Proca configurations presented in \cite{Ma2024}, and Dirac stars in anti-de Sitter spacetime investigated in \cite{Zhang2024} which further extended theoretical description of gravitationally bound spinor systems.
An additional degree of complexity arises when nonlinear interactions are introduced into spinor sector, however early investigations of nonlinear spinor field theories by Finkelstein and his collaborators in \cite{Finkelstein1951}, together with nonlinear classical spinor model developed by Soler and presented in \cite{Soler1970}, provided theoretical foundations for exploring self-interacting fermionic configurations. Within a gravitational setting, Dzhunushaliev and his co-author Folomeev \cite{Dzhunushaliev2019a} demonstrated that nonlinear spinor self interactions can substantially modify masses, characteristic radii, as well as equilibrium properties of Dirac stars, while their subsequent extensions incorporated Maxwell along with Proca fields \cite{Dzhunushaliev2019b}, also non Abelian Yang-Mills interactions studied in \cite{Dzhunushaliev2020}, thereby revealing sensitivity of self-gravitating spinor systems to additional field degrees of freedom.
Diversity of self-gravitating fermionic configurations further demonstrated through Dirac-boson stars
which presented in \cite{Liang2023}, $\kappa$-Dirac stars studied in \cite{Sun2024Kappa}, and minimal geometric deformation of Yang-Mills-Dirac stellar configurations examined in \cite{DaRocha2020}. 
More recently, Alibekova and his collaborators \cite{Alibekova2026} examined Dirac stars within quadratic $F(R)=R+\alpha R^2$ gravity and investigated how additional gravitational degree of freedom influences their mass spectrum and characteristic frequencies. 

Inclusion of electric charge introduces a further competition between gravitational binding as well as electromagnetic repulsion, however charged self-gravitating field configurations attracted considerable attention, particularly in context of charged boson stars, whose properties revisited in \cite{Lopez2023}. For fermionic configurations, Finster et al. \cite{Finster1999b} constructed particle-like solutions of coupled Einstein-Dirac-Maxwell equations, establishing existence of electrically charged spinor configurations in Einstein's general relativity. Relationship between rotating gauged boson stars as well as their fermionic counterparts subsequently investigated in~\cite{Herdeiro2022}, however recent analysis by Hern\'andez-M\'arquez and his collaborator Alcubierre \cite{Hernandez2026} explored binding energy, compactness, along with equilibrium families of charged Dirac stars, emphasizing role of charge to mass ratio in determining whether gravitationally bound configurations can exist. Most conventional investigations of charged Dirac stars employ Maxwell electrodynamics, for which electromagnetic field responds linearly to its source, however, nonlinear electromagnetic interactions may alter this balance by modifying relation between electric field as well as enclosed charge  \cite{Bajardi:2026fwl, Capozziello:2023vvr, Capozziello:2019uvk}. Gravitational consequences of nonlinear electrodynamics explored extensively in connection with regular black hole geometries, including models developed in \cite{AyonBeato1998} and Bronnikov presented in \cite{Bronnikov2001}, while related developments extended nonlinear electromagnetic scenario to self-gravitating spinor configurations. Huang and his collaborators \cite{Huang2025} investigated Dirac stars associated with a Bardeen-type magnetic background and obtained frozen star configurations, while Sun and his collaborators \cite{Sun2024} considered charged Dirac fields in a related nonlinear electromagnetic setting. Particularly relevant scenario is provided by EH nonlinear electrodynamics, which describes effective electromagnetic response of quantum vacuum, while pioneering analysis of Heisenberg and his collaborator Euler \cite{Heisenberg1936} established that virtual charged particle fluctuations generate nonlinear corrections to classical Maxwell theory. Schwinger \cite{Schwinger1951} subsequently formulated effective action in a gauge invariant quantum field theoretic idea, while its physical foundations, mathematical structure, also subsequent developments reviewed by Dunne in 2012 \cite{Dunne2012}, however in weak field expansion, EH effective Lagrangian contains quartic electromagnetic corrections that encode photon photon interactions induced by vacuum polarization. These nonlinear effects acquired considerable phenomenological interest through evidence for light by light scattering reported by ATLAS Collaboration in 2017 and presented in \cite{ATLAS2017}, as well as observational evidence consistent with vacuum birefringence in isolated neutron star RX J1856.5$-$3754 \cite{Mignani2017}. Relevance of strong electromagnetic fields as well as associated pair production phenomena in relativistic astrophysical environments also discussed extensively \cite{Ruffini2010}. Coupling of EH electrodynamics to general relativity introduces modifications to both electromagnetic stress energy tensor and resulting spacetime geometry, in particular, Yajima and his collaborator Tamaki \cite{Yajima2001} investigated black hole solutions in EH theory as well as demonstrated that nonlinear electromagnetic corrections influence corresponding gravitational configuration. Gravitational implications of EH electrodynamics also investigated through exact gravitational lensing relations for static charged black holes \cite{Amaro2022}, furthermore, Mushtaq and his collaborators \cite{Mushtaq2024} examined weak gravitational lensing as well as fundamental frequencies in Einstein-EH black hole spacetimes, illustrating influence of nonlinear electromagnetic corrections on their optical and dynamical properties. More recently, charged anisotropic compact stars within same nonlinear electromagnetic scenario and investigated resulting modifications to their mass radius properties presented in \cite{Ditta2026,Ditta2026a}. 

Despite substantial progress in study of charged Dirac stars and nonlinear electromagnetic compact objects, combined influence of EH nonlinear electrodynamics on self-gravitating charged spinor configurations remains insufficiently explored within scenario considered here, in particular, it is important to determine how nonlinear constitutive relation between electric field as well as its source modifies distribution of electromagnetic energy, equilibrium spinor profiles, along with global properties of resulting configurations. Related issue concerns interpretation of changes in maximal stellar mass, since total electric charge is directly related to conserved fermion number, comparing solutions at a fixed central spinor amplitude does not necessarily isolate intrinsic effect of nonlinear electrodynamics. Motivated by these considerations, present work investigates static, spherically symmetric Dirac stars in Einstein gravity coupled to EH nonlinear electrodynamics, with an additional quartic spinor self interaction, while electromagnetic sector is described by EH Lagrangian truncated at fourth order in field strength, whereas spinor field is minimally coupled to an Abelian $U(1)$ gauge potential. Employing a two spinor ansatz compatible with spherical symmetry, we derive coupled Einstein Dirac EH field equations as well as formulate corresponding dimensionless boundary value problem, while regular and asymptotically flat configurations are obtained numerically by determining allowed spinor eigenfrequencies. Resulting solutions are examined by varying electric charge coupling, nonlinear electromagnetic coupling, as well as spinor self interaction parameter however uncharged and Maxwell limits provide consistency checks against established Dirac star configurations \cite{Herdeiro2017,Finster1999b,Alibekova2026,Hernandez2026}. Present study examines how EH nonlinear electrodynamics modifies structure as well as equilibrium of charged Dirac stars, unlike Maxwell theory, nonlinear relation between electric displacement as well as field strength suppresses electric field for a given enclosed charge, thereby influencing electromagnetic repulsion, gravitational mass, along with spinor distribution. We investigate resulting changes in metric as well as matter profiles, pressure anisotropy, energy conditions, binding energy, compactness, and mass frequency relations, however our results reveal that, at fixed fermion number, EH corrections produce only minor changes in stellar mass while enhancing compactness and gravitational binding. Maximum mass decreases because nonlinear electromagnetic response shifts equilibrium turning point toward smaller particle numbers, in particular, for $\bar e=0.6$, increasing $\bar\alpha$ from $0$ to $200$ reduces maximum mass from $0.924$ to $0.840$, corresponding to an approximately $9\%$ reduction. Importantly, physical interpretation of these effects depends on validity of quartic EH approximation, while in controlled weak field QED regime, relative mass correction remains extremely small, scaling as $(E_{\max}/E_c)^2$, whereas larger modifications at strong coupling characterize quartic theory as an effective nonlinear electromagnetic model. Our paper is organized as follows, in section~\ref{sec:model}, we introduce Einstein Dirac EH scenario, derive governing field equations, specify boundary conditions, as well as describe numerical construction of regular configurations. Section~\ref{sec:results} presents numerical results as well as discusses influence of charge, nonlinear electromagnetic coupling, and spinor self interaction on stellar structure, equilibrium conditions, along with global mass spectrum, however section~\ref{sec:conclusions} summarizes principal findings and outlines possible directions for further investigation.

\section{Model and field equations}\label{sec:model}

We use the metric signature $(+,-,-,-)$, natural units $c=\hbar=1$ and Heaviside-Lorentz units for the electromagnetic field. The action is
\begin{equation}
S=\int d^4x\sqrt{-g}\left[-\frac{R}{16\pi G}+\mathcal L_{\rm sp}+\mathcal L_{\rm EH}\right].
\label{eq:action}
\end{equation}
The electromagnetic part is the EH Lagrangian truncated at quartic order,
\begin{equation}
\mathcal L_{\rm EH}=-\frac14\mathcal F+\alpha\mathcal F^2,\qquad \mathcal F=F_{\mu\nu}F^{\mu\nu}.
\end{equation}
The term with $\mathcal G=F_{\mu\nu}\tilde F^{\mu\nu}$ is omitted because it vanishes for the purely electric fields considered here. The spinor Lagrangian is
\begin{eqnarray}
\mathcal L_{\rm sp}&=&\frac{i}{2}\left(\bar\psi\gamma^\mu D_\mu\psi-\overline{D_\mu\psi}\,\gamma^\mu\psi\right)\nonumber\\
&&-\mu\bar\psi\psi+\frac{\lambda}{2}\left(\bar\psi\psi\right)^2,
\end{eqnarray}
where $D_\mu=\nabla_\mu+ieA_\mu$ is the gauge covariant derivative, $\nabla_\mu$ contains the spin connection, $\mu$ is the mass of the fermion, $e$ its gauge coupling and $\lambda$ the self-interaction constant. For $\lambda>0$ the self-interaction is attractive \cite{Dzhunushaliev2019a}. Varying the action gives
\begin{align}
&G_\mu^{\ \nu}=8\pi G\,[T_\mu^{\ \nu({\rm sp})}+T_\mu^{\ \nu({\rm EH})}],\label{eq:einstein}\\
&i\gamma^\mu D_\mu\psi-\mu\psi+\lambda(\bar\psi\psi)\psi=0,\label{eq:dirac}\\
&\nabla_\mu\left[\left(1-8\alpha\mathcal F\right)F^{\mu\nu}\right]=e\,\bar\psi\gamma^\nu\psi .\label{eq:maxwell}
\end{align}
The spinor stress tensor is the symmetrized expression of Ref.~\cite{Alibekova2026} with $\nabla_\mu$ replaced by $D_\mu$, and the stress tensor of the EH field is $T_{\mu\nu}^{({\rm EH})}=-4\mathcal L_{\mathcal F}F_{\mu\lambda}F_\nu^{\ \lambda}+g_{\mu\nu}\mathcal L_{\rm EH}$ with $\mathcal L_{\mathcal F}=\partial\mathcal L_{\rm EH}/\partial\mathcal F$. Both stress tensors follow from the variation of the matter part of Eq.~\eqref{eq:action} and enter Eq.~\eqref{eq:einstein} with the same coupling $8\pi G$. In the dimensionless variables introduced below this common factor is absorbed into the definitions of $m$, $f$, $g$ and $E$, which is why it does not appear explicitly in Eq.~\eqref{eq:m}.
For static and spherically symmetric configurations we write
\begin{eqnarray}
ds^2&=&N\sigma^2dt^2-\frac{dr^2}{N}-r^2\left(d\theta^2+\sin^2\theta\,d\varphi^2\right),\nonumber\\
N&=&1-\frac{2Gm(r)}{r},\qquad A_\mu dx^\mu=V(r)\,dt ,
\end{eqnarray}
where $m(r)$ is the mass inside the radius $r$. For the spinor we use the two-fermion ansatz of Refs.~\cite{Herdeiro2017,Alibekova2026}, which contains two real radial functions $f(r)$, $g(r)$ and the frequency $\Omega$. It gives a vanishing total spin and a spherically symmetric stress tensor. The only nonzero component of the field tensor is $F_{rt}=V'$. In the local orthonormal frame the electric field is $E=V'/\sigma$ (up to sign) and $\mathcal F=-2E^2$. The time component of Eq.~\eqref{eq:maxwell} then becomes the Gauss law
\begin{equation}
\left(r^2D\right)'=\frac{e\,r^2\left(f^2+g^2\right)}{\sqrt N},\quad D=E\left(1+16\alpha E^2\right),
\label{eq:gauss}
\end{equation}
where we used $\psi^\dagger\psi=f^2+g^2$. The quantity $Q(r)=r^2D$ is the charge inside the radius $r$ divided by $4\pi$. The displacement $D$ is fixed by the charge, while the field $E$ is obtained from the nonlinear relation in Eq.~\eqref{eq:gauss}. Since the gauge field enters the Dirac equation only through $D_t$, its effect on the spinor is to replace the frequency $\Omega$ by the local value
\begin{equation}
\omega(r)=\Omega-eV(r).
\end{equation}
For a radial electric field the EH stress tensor has $T_0^{\ 0}=T_1^{\ 1}$, and the energy density and the pressures are
\begin{eqnarray}
\rho_{\rm EH}&=&ED-\mathcal L_{\rm EH}=\frac{E^2}{2}+12\alpha E^4,\nonumber\\
p_r^{\rm EH}&=&-\rho_{\rm EH},\qquad p_t^{\rm EH}=E^2/2+4\alpha E^4 .
\label{eq:EHstress}
\end{eqnarray}
These follow from $T_\mu^{\ \nu({\rm EH})}$ with $\mathcal F=-2E^2$ and $\mathcal L_{\mathcal F}=-1/4-4\alpha E^2$, and they are used without change in Eqs.~\eqref{eq:m} and \eqref{eq:pressures}. The field contributes a negative radial pressure (a tension along the field lines) and a positive tangential pressure. Because $\rho_{\rm EH}+p_r^{\rm EH}=0$, the field does not enter the equation for $\sigma$.

We introduce dimensionless quantities using the Compton wavelength $\lambda_c=1/\mu$:
\begin{eqnarray}
&&x=\mu r,\quad \bar\Omega=\frac{\Omega}{\mu},\quad \bar m=\frac{\mu m}{M_{\rm Pl}^2},\quad \bar\lambda=\frac{\lambda}{4\pi G},\nonumber\\
&&(\bar f,\bar g)=\sqrt{4\pi}\,\lambda_c^{3/2}\frac{\mu}{M_{\rm Pl}}(f,g),\quad \bar E=\frac{\sqrt{4\pi}\,E}{M_{\rm Pl}\,\mu},\nonumber\\
&&\bar e=\frac{e\,M_{\rm Pl}}{\sqrt{4\pi}\,\mu},\quad \bar\alpha=\frac{\alpha M_{\rm Pl}^2\mu^2}{4\pi},\nonumber\\
&&\phi=\frac{eV}{\mu},\quad q=\frac{\sqrt{4\pi}\,\mu\,Q}{M_{\rm Pl}} .
\end{eqnarray}
The parameter $\bar e$ measures the electric repulsion against the gravitational attraction. A Reissner-Nordstr\"om exterior with $q=M$ is extremal, and for weakly bound matter this corresponds to $\bar e=1$. Dropping the bars, with $n=f^2+g^2$ and $P=f^2-g^2$, the field equations become
\begin{eqnarray}
&&f'+f\left[\frac{N'}{4N}+\frac{\sigma'}{2\sigma}+\frac1x\left(1+\frac1{\sqrt N}\right)\right]\nonumber\\
&&\qquad+g\left(\frac1{\sqrt N}-\frac{\omega}{N\sigma}+\frac{\lambda P}{\sqrt N}\right)=0,\label{eq:f}\\
&&g'+g\left[\frac{N'}{4N}+\frac{\sigma'}{2\sigma}+\frac1x\left(1-\frac1{\sqrt N}\right)\right]\nonumber\\
&&\qquad+f\left(\frac1{\sqrt N}+\frac{\omega}{N\sigma}+\frac{\lambda P}{\sqrt N}\right)=0,\label{eq:g}\\
&&m'=x^2\left[\frac{\omega\,n}{\sqrt N\sigma}+\frac\lambda2P^2+\frac{E^2}{2}+12\alpha E^4\right],\label{eq:m}\\
&&\sigma'=x\sigma\left[\frac{\omega\,n}{N^{3/2}\sigma}-\frac{fg'-gf'}{\sqrt N}\right],\label{eq:sigma}\\
&&q'=\frac{e\,x^2n}{\sqrt N},\qquad \phi'=-e\,\sigma E,\label{eq:qphi}\\
&&E\left(1+16\alpha E^2\right)=\frac{q}{x^2},\label{eq:em}
\end{eqnarray}
with $\omega=\Omega-\phi$ and $N=1-2m/x$. For $e=0$ Eqs.~\eqref{eq:f}-\eqref{eq:sigma} reduce to the Einstein-Dirac equations of Refs.~\cite{Herdeiro2017,Alibekova2026}. For $\alpha\ge0$ the cubic relation \eqref{eq:em} has one real root,
\begin{equation}
E=\frac{1}{2\sqrt{3\alpha}}\sinh\left[\frac13\,{\rm arcsinh}\left(6\sqrt{3\alpha}\,\frac{q}{x^2}\right)\right].
\label{eq:Eroot}
\end{equation}
For weak fields this gives the Coulomb law $E\simeq q/x^2$. For strong fields it gives $E\simeq(q/16\alpha x^2)^{1/3}$, so the field grows much more slowly than the enclosed charge. This saturation of the field controls most of the EH effects found below.

Regularity at the center requires
\begin{eqnarray}
f&\simeq&\frac{g_c}{3}\left(\frac{\omega_c}{\sigma_c}-1+\lambda g_c^2\right)x,\qquad g\simeq g_c,\nonumber\\
m&\simeq&\frac{x^3}{3}\left(\frac{\omega_c g_c^2}{\sigma_c}+\frac\lambda2g_c^4\right),\qquad q\simeq\frac{e\,g_c^2x^3}{3},
\end{eqnarray}
so that the electric field vanishes linearly at the center, $E\simeq e\,g_c^2x/3$. Far from the star the spinor decays as $f,g\sim\exp(-\sqrt{1-\Omega^2}\,x)$, and the metric approaches the EH-corrected Reissner-Nordstr\"om form \cite{Yajima2001}
\begin{eqnarray}
N&\simeq&1-\frac{2M}{x}+\frac{q_\infty^2}{x^2}-\frac{8\alpha}{5}\frac{q_\infty^4}{x^6},\nonumber\\
\phi&\simeq&\frac{e\,q_\infty}{x}-\frac{16\alpha e\,q_\infty^3}{5x^5},
\end{eqnarray}
with $M$ the total mass and $q_\infty$ the total charge. Localized solutions exist only for $\Omega<1$. The total charge equals $e$ times the number of particles,
\begin{equation}
q_\infty=e\,\mathcal N,\qquad \mathcal N=\int_0^\infty\frac{x^2n}{\sqrt N}\,dx,
\end{equation}
and the binding energy is $E_b=M-\mathcal N$ in units of $M_{\rm Pl}^2/\mu$. A negative $E_b$ means that the star has less energy than the same number of free fermions.

The total radial pressure follows from Eq.~\eqref{eq:sigma}, which reads $N\sigma'/(x\sigma)=\rho+p_r$. The tangential pressure follows from the conservation law $\nabla_\mu T^\mu_{\ r}=0$:
\begin{eqnarray}
p_r&=&-\sqrt N\left(fg'-gf'\right)-\frac{\lambda}{2}P^2-\rho_{\rm EH},\nonumber\\
p_t&=&p_r+\frac x2\left[p_r'+(\rho+p_r)\left(\frac{N'}{2N}+\frac{\sigma'}{\sigma}\right)\right],
\label{eq:pressures}
\end{eqnarray}
where $\rho=m'/x^2$ is the total energy density. The first relation shows the three contributions to the radial stress: a kinetic part of the spinor, a negative part $-\lambda P^2/2$ from the self-interaction, and the field tension $-\rho_{\rm EH}$. The second relation follows from the radial component of the conservation law and gives the tangential pressure once $p_r$, $\rho$ and the metric functions are known.

The system is invariant under $\sigma\to k\sigma$, $\phi\to k\phi$, $\Omega\to k\Omega$, and $\Omega$ and $\phi$ appear only through $\omega$. We therefore integrate with $\sigma(0)=1$ and $\phi(0)=0$ and afterwards rescale to $\sigma(\infty)=1$, $\phi(\infty)=0$. For given $g_c$, $e$, $\alpha$ and $\lambda$, regular and localized solutions exist only for discrete values of $\Omega$. We find the nodeless solution by shooting in $\Omega$ with bisection and refine it with a collocation boundary value solver \cite{Virtanen2020}, with $\Omega$ as an unknown parameter. Families are followed by continuation in $g_c$, and by pseudo-arclength continuation in $(g_c,\Omega)$ where the mass varies too steeply with $g_c$. The solver settings, the boundary conditions and convergence tests are given in Appendix~\ref{app:numerics}, and the expansion near the center is derived in Appendix~\ref{app:center}. We checked the code in four ways. For $e=0$ it gives $\Omega=0.769587$ and $\sigma_c=0.509157$ at $g_c=0.5$, in full agreement with Ref.~\cite{Alibekova2026}. The relation $q_\infty=e\mathcal N$ holds to five digits. The mass $m(\infty)$ agrees with $\int_0^\infty x^2\rho\,dx$ to six digits. Finally, shooting and collocation give the same frequency to $10^{-6}$.

It is useful to relate $\bar\alpha$ to particle physics. In weak fields the EH Lagrangian of a charged fermion of mass $m_\ell$ and fine structure constant $\alpha_f$ is $\frac12(E^2-B^2)+\frac{2\alpha_f^2}{45m_\ell^4}(E^2-B^2)^2+\dots$ \cite{Heisenberg1936,Dunne2012}, which corresponds to $\alpha=\alpha_f^2/(90m_\ell^4)$. With the definitions above,
\begin{equation}
\bar\alpha=\frac{\alpha_f\,\bar e^{\,2}}{360\pi}\left(\frac{\mu}{m_\ell}\right)^4 .
\end{equation}
If the loop particle were the star fermion itself ($m_\ell=\mu$), $\bar\alpha$ would be of order $10^{-5}$. Larger values require a lighter charged particle with $m_\ell\ll\mu$ coupled to the same U(1) field.

The quartic Lagrangian is the weak-field expansion of the full EH action. The expansion parameter is $E/E_c$, where $E_c=m_\ell^2/e$ is the critical field of the same charged particle of mass $m_\ell$ that circulates in the loop and produces $\alpha$ \cite{Dunne2012,Schwinger1951}. We assume that this particle couples to the same U(1) field as the star fermion with the same charge $e$. Using $\alpha=\alpha_f^2/(90m_\ell^4)$ and $e^2=4\pi\alpha_f$, the dimensionless combination that measures the nonlinearity becomes
\begin{equation}
16\bar\alpha\bar E^2=16\alpha E^2=\frac{2\alpha_f}{45\pi}\left(\frac{E}{E_c}\right)^2 .
\label{eq:validity}
\end{equation}
The condition $E\lesssim E_c$ is therefore $16\bar\alpha\bar E^2\lesssim2\alpha_f/(45\pi)\approx1.0\times10^{-4}$ for $\alpha_f=1/137$. This condition sets the scale of the nonlinearity but is not an accuracy criterion by itself. The terms neglected in the truncation are smaller than the quartic term by further powers of $(E/E_c)^2$ \cite{Dunne2012}, so the quartic Lagrangian is quantitatively accurate only for $E\ll E_c$. In this controlled regime the EH term is a small perturbation of the Maxwell theory. When $16\bar\alpha\bar E^2$ is of order one, Eq.~\eqref{eq:validity} gives $E\sim10^2E_c$, where higher order terms of the effective action and pair creation \cite{Schwinger1951,Ruffini2010} are no longer negligible. We therefore separate two regimes. For $E_{\max}\ll E_c$, that is $16\bar\alpha\bar E_{\max}^2\ll10^{-4}$, our results are controlled QED corrections. For larger values they describe the quartic Lagrangian as an effective nonlinear electrodynamics with a free coupling, in the same spirit as the EH black hole and compact star models of Refs.~\cite{Yajima2001,Ditta2026}. The value of $16\bar\alpha\bar E_{\max}^2$ is listed for every solution in Table~\ref{tab:gc05}.

\section{Results and discussion}\label{sec:results}

\begin{table*}[!t]
\centering
\caption{Solutions with $g_c=0.5$. $R_{99}$ contains $99\%$ of the particles and $R_M$ contains $99\%$ of the total mass. $E_{\max}$ is the largest electric field and $16\bar\alpha E_{\max}^2$ measures the strength of the EH nonlinearity (the QED weak-field regime is $16\bar\alpha E_{\max}^2\lesssim10^{-4}$).}
\label{tab:gc05}
\begin{tabular}{|c|c|c|c|c|c|c|c|c|c|c|c|c|c|c|}
\hline
$\lambda$ & $\bar e$ & $\bar\alpha$ & $\Omega$ & $\sigma_c$ & $M$ & $q_\infty$ & $E_b$ & $R_{99}$ & $R_M$ & $E_{\max}$ & $16\bar\alpha E_{\max}^2$\\
\hline
0 & 0 & $\cdot$ & 0.769587 & 0.509157 & 0.67381 & 0 & $-0.01845$ & 5.424 & 5.2 & 0 & 0\\
\hline
0 & 0.3 & 50 & 0.782345 & 0.500889 & 0.70282 & 0.21557 & $-0.01574$ & 5.490 & 6.0 & 0.0204 & 0.33\\
\hline
0 & 0.6 & 50 & 0.822889 & 0.481271 & 0.78129 & 0.47045 & $-0.00279$ & 5.689 & 14.2 & 0.0315 & 0.79\\
\hline
0 & 0.75 & 50 & 0.854564 & 0.467730 & 0.84155 & 0.62376 & $\phantom{-}0.00987$ & 5.854 & 23.1 & 0.0357 & 1.02\\
\hline
0 & 0.6 & 0 & 0.822725 & 0.459934 & 0.84453 & 0.51240 & $-0.00947$ & 5.751 & 15.6 & 0.0587 & 0\\
\hline
0 & 0.6 & 10 & 0.824087 & 0.470936 & 0.81279 & 0.49091 & $-0.00539$ & 5.736 & 14.8 & 0.0440 & 0.31\\
\hline
0 & 0.6 & 200 & 0.819153 & 0.489606 & 0.75446 & 0.45387 & $-0.00199$ & 5.629 & 13.6 & 0.0219 & 1.54\\
\hline
1 & 0.6 & 50 & 0.828369 & 0.530314 & 0.69536 & 0.41769 & $-0.00079$ & 5.641 & 12.6 & 0.0307 & 0.76\\
\hline
10 & 0.6 & 50 & 0.539037 & 0.385263 & 0.62885 & 0.46769 & $-0.15063$ & 2.826 & 17.4 & 0.0489 & 1.91\\
\hline
20 & 0.6 & 50 & 0.586621 & 0.239976 & 0.39906 & 0.23033 & $\phantom{-}0.01519$ & 2.049 & 6.6 & 0.0590 & 2.79\\
\hline
\end{tabular}
\end{table*}

\begin{figure*}[!t]\centering
\includegraphics[width=\textwidth]{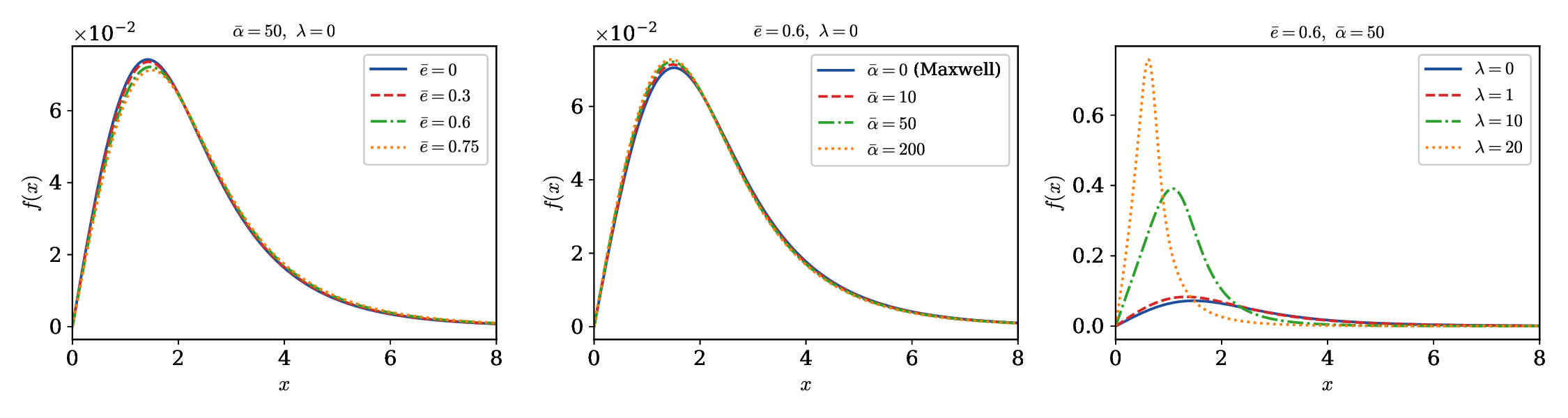}\\
\includegraphics[width=\textwidth]{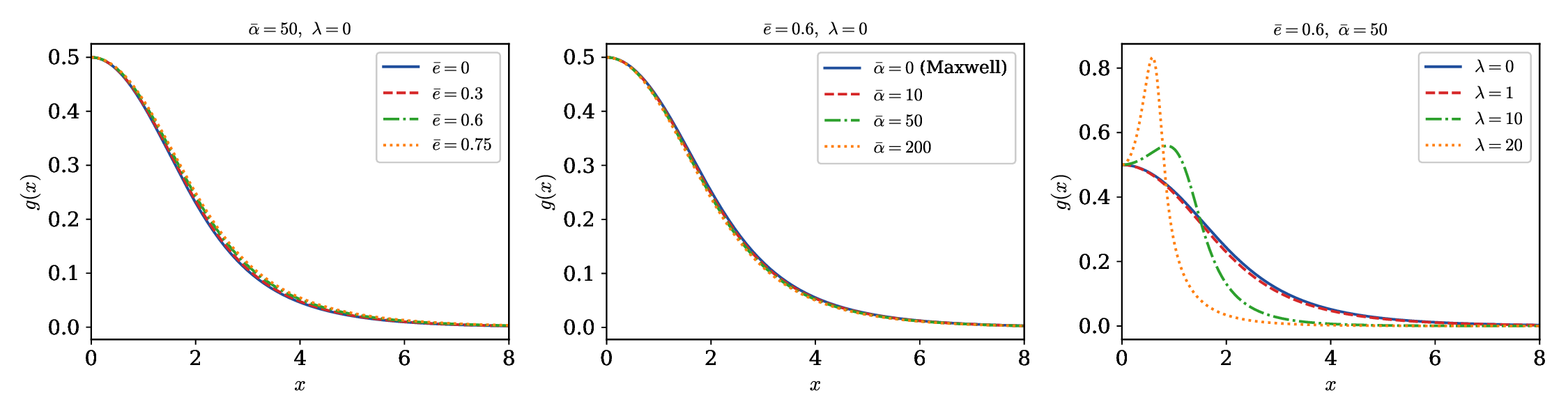}
\caption{Spinor functions $f(x)$ (top) and $g(x)$ (bottom) for different charge (left), EH coupling (middle) and self-interaction (right).}
\label{fig:fg}
\end{figure*}

\begin{figure*}[!t]\centering
\includegraphics[width=\textwidth]{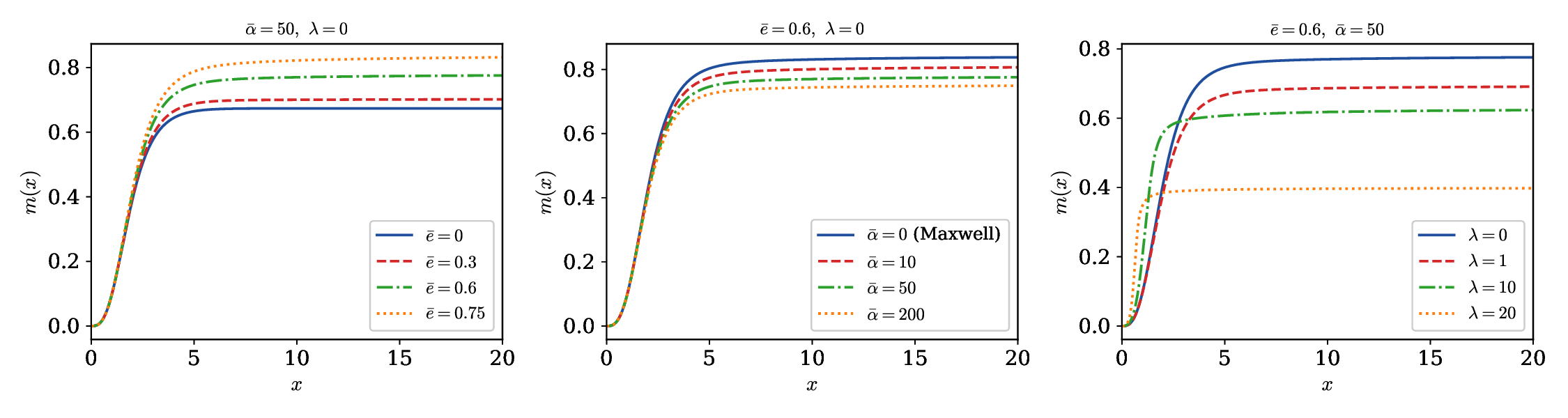}\\
\includegraphics[width=\textwidth]{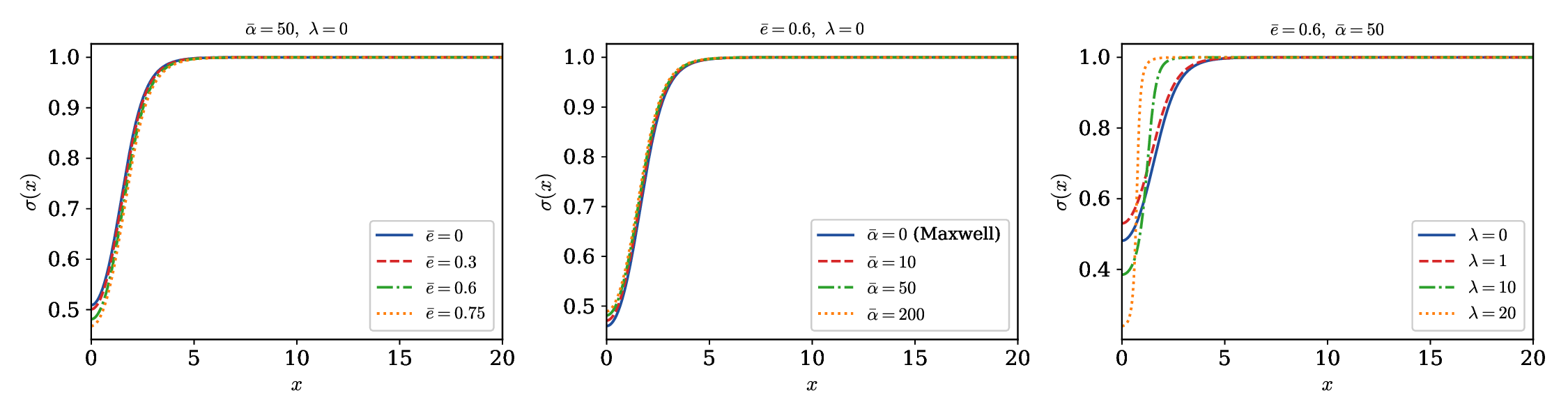}
\caption{Mass function $m(x)$ (top) and metric function $\sigma(x)$ (bottom).}
\label{fig:metric}
\end{figure*}

\begin{figure*}[!t]\centering
\includegraphics[width=\textwidth]{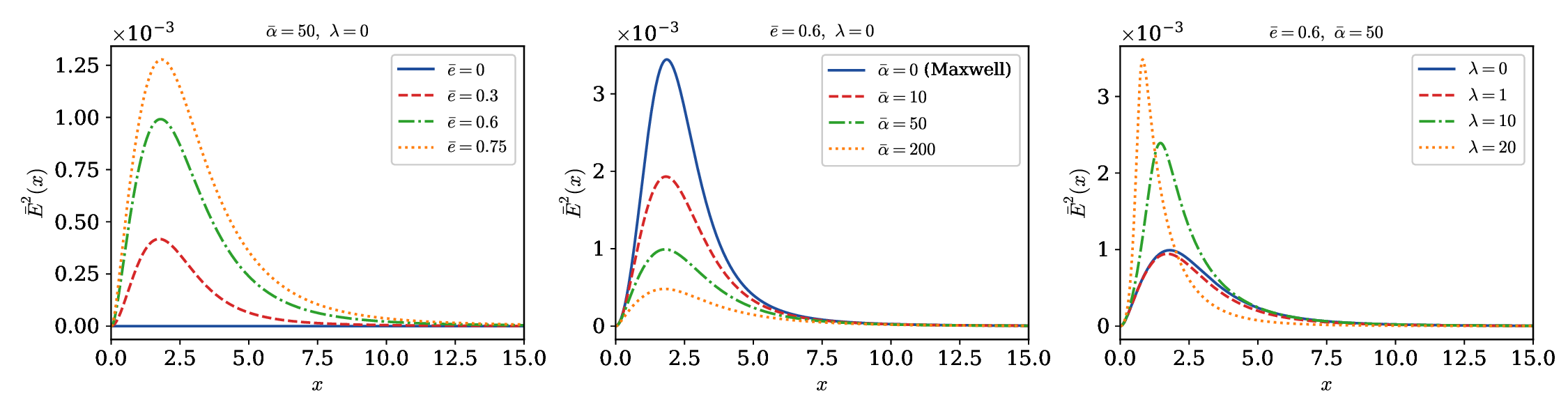}\\
\includegraphics[width=\textwidth]{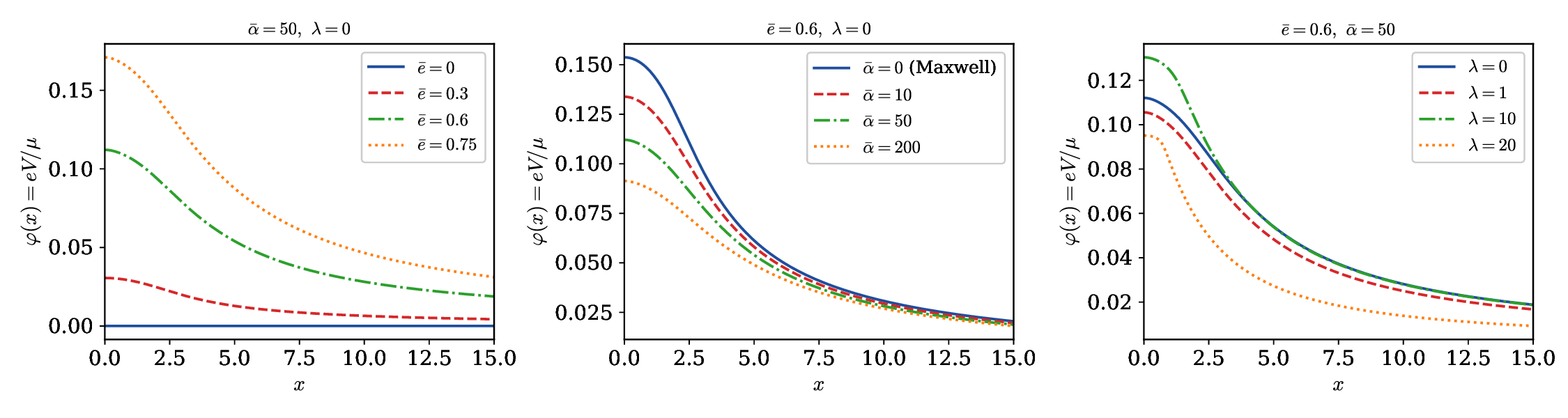}
\caption{Electric field intensity $E^2$ (top) and electric potential $\phi=eV/\mu$ (bottom).}
\label{fig:field}
\end{figure*}

\begin{figure*}[!t]\centering
\includegraphics[width=\textwidth]{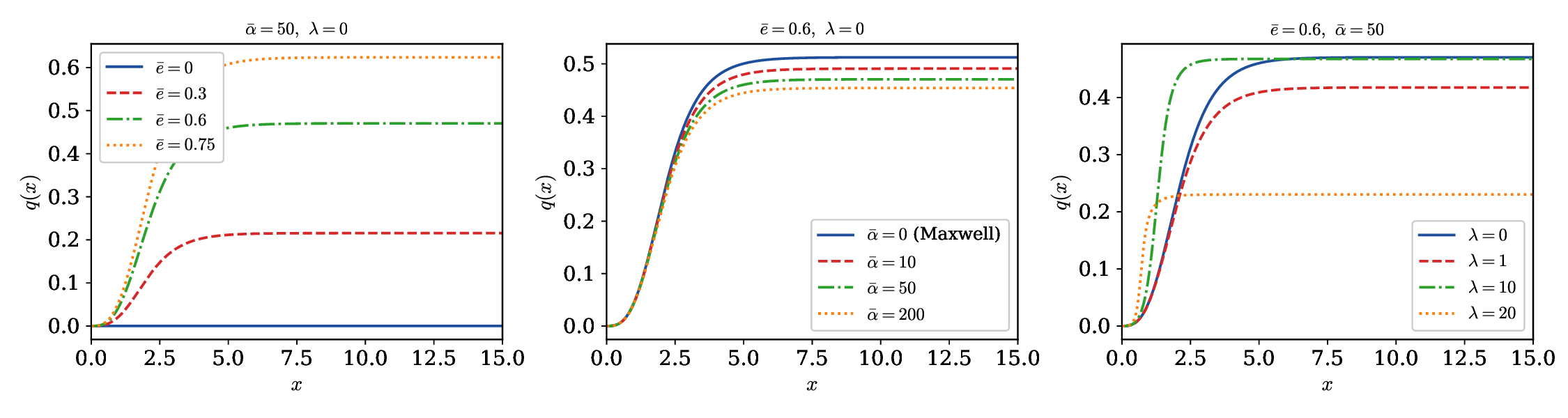}\\
\includegraphics[width=\textwidth]{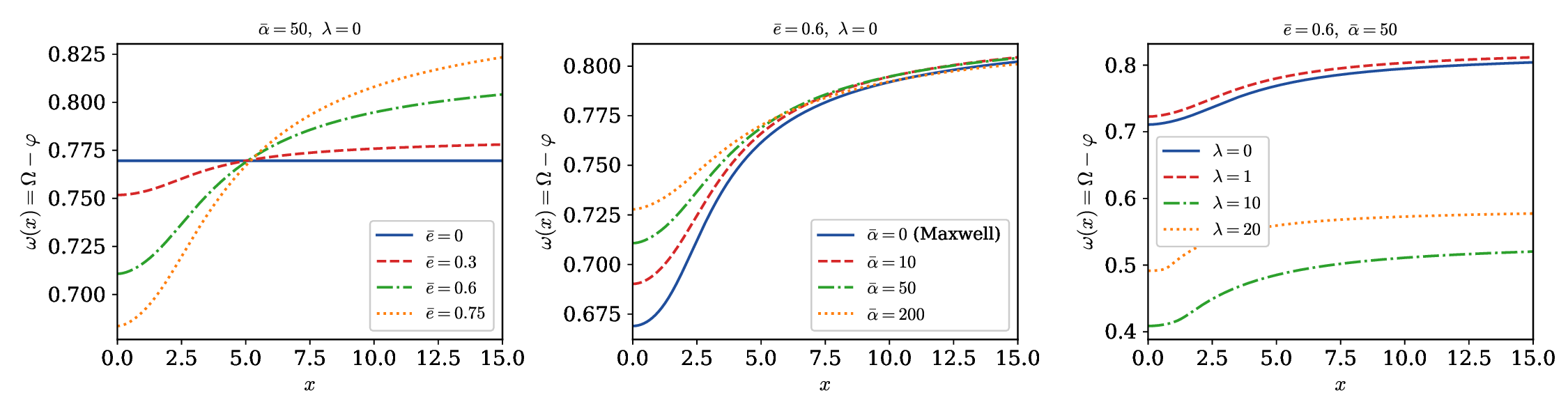}
\caption{Enclosed charge $q(x)$ (top) and local frequency $\omega(x)$ (bottom).}
\label{fig:charge}
\end{figure*}

\begin{figure*}[!t]\centering
\includegraphics[width=\textwidth]{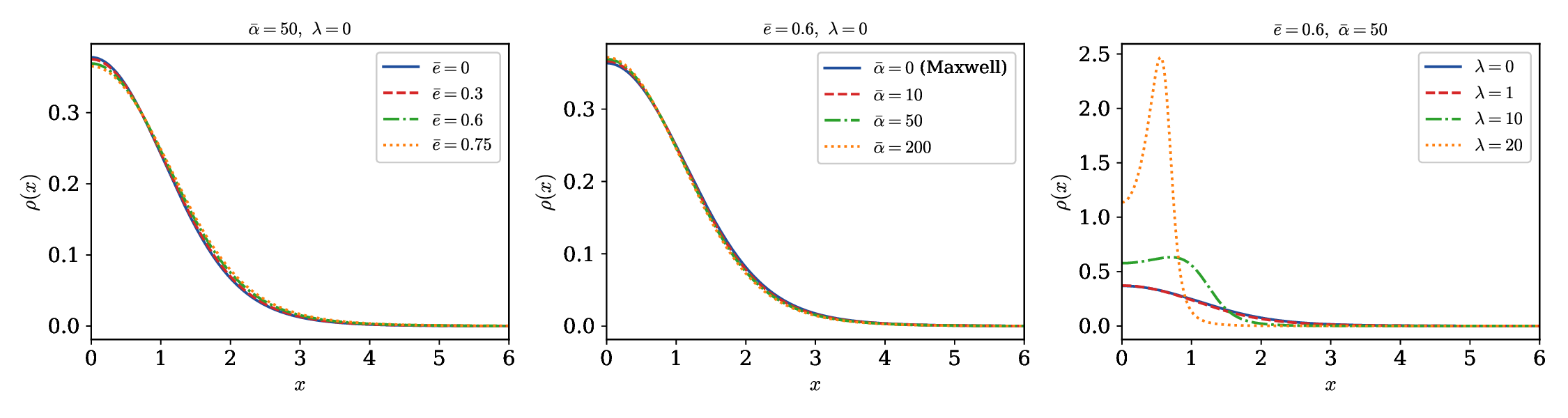}
\caption{Total energy density $\rho(x)$ for different charge (left), EH coupling (middle) and self-interaction (right).}
\label{fig:stress}
\end{figure*}

\begin{figure*}[!t]\centering
\includegraphics[width=\textwidth]{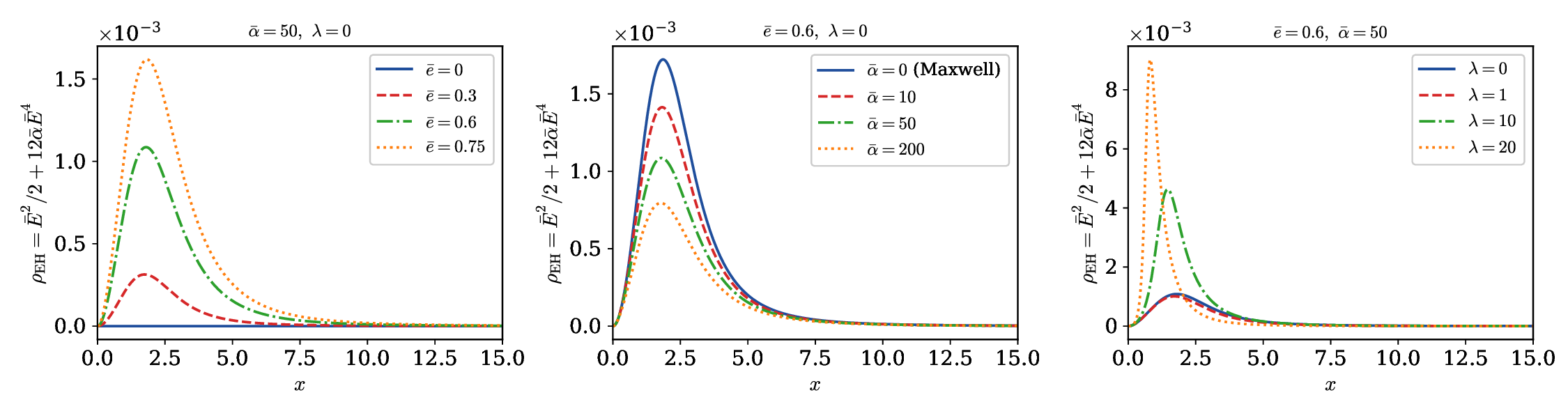}\\
\includegraphics[width=0.68\textwidth]{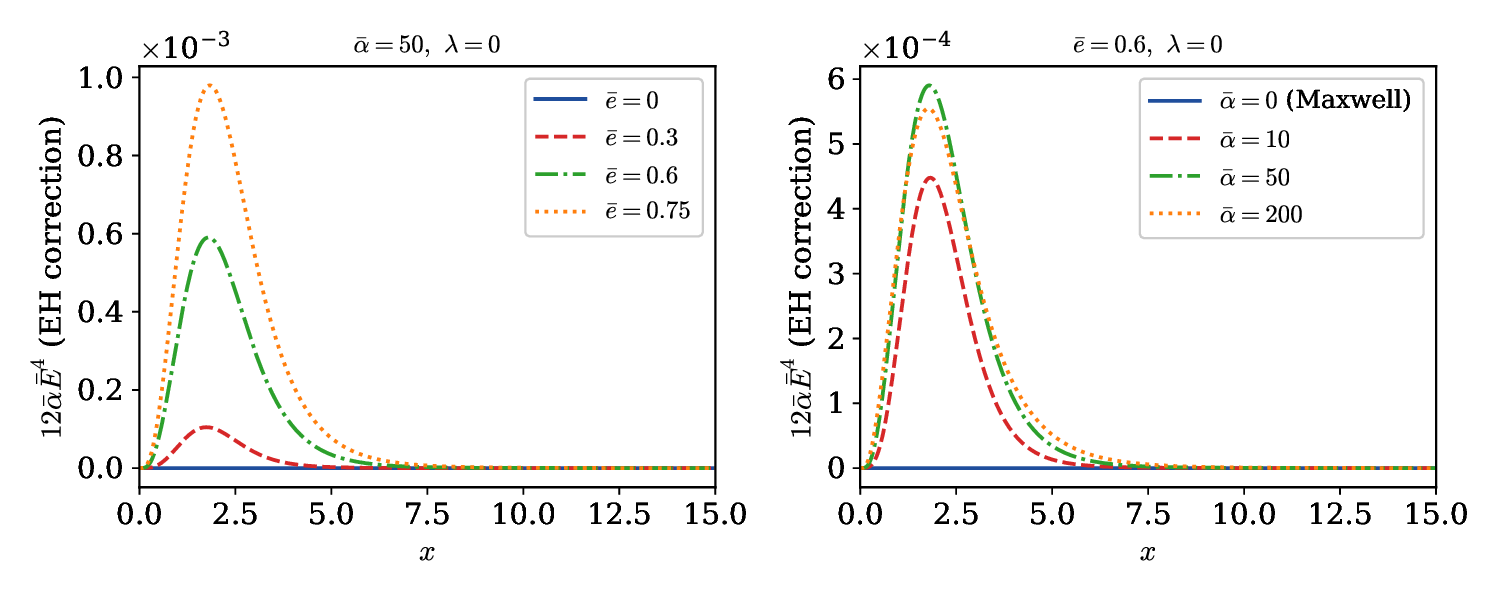}
\caption{Electromagnetic energy density $\rho_{\rm EH}$ (top) and its EH part $12\alpha E^4$ (bottom).}
\label{fig:aniso}
\end{figure*}

\begin{figure*}[!t]\centering
\includegraphics[width=\textwidth]{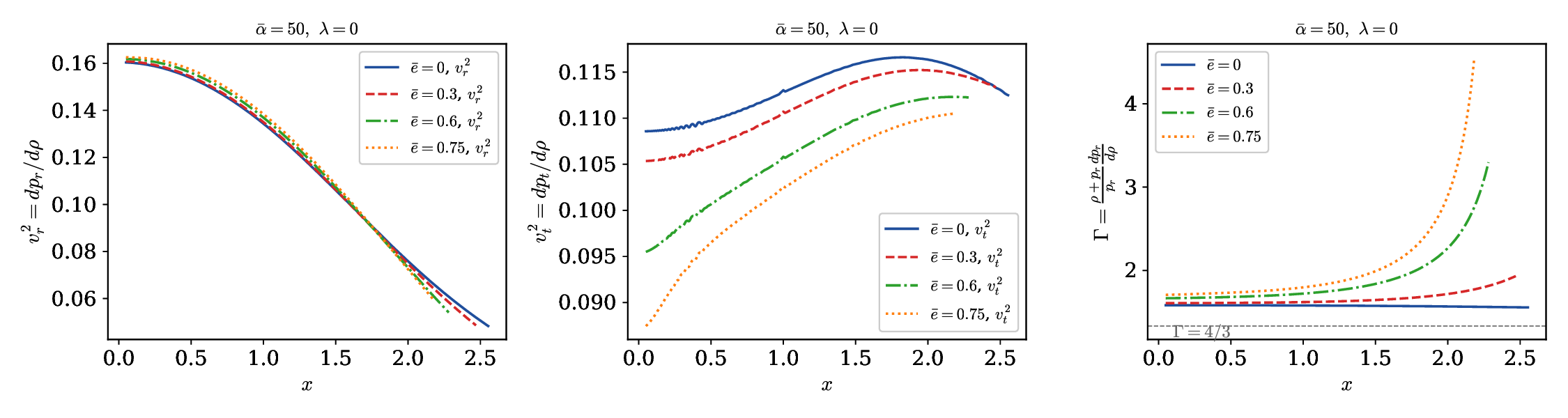}
\caption{Radial sound speed (left), tangential sound speed (middle) and adiabatic index (right) in the core of the stars.}
\label{fig:sound}
\end{figure*}

\begin{figure*}[!t]\centering
\includegraphics[width=\textwidth]{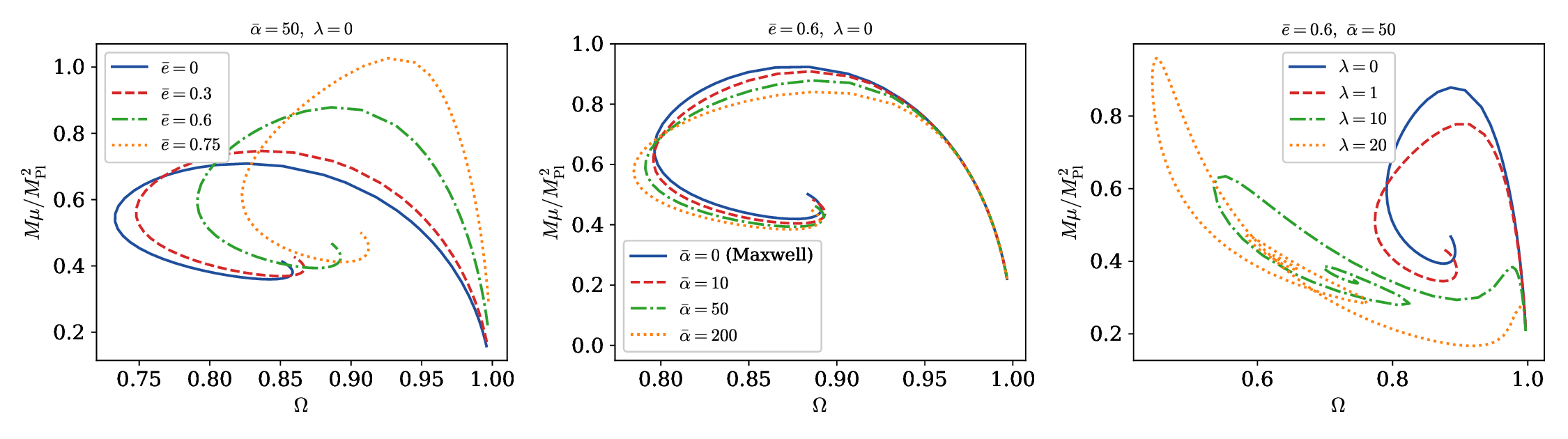}\\
\includegraphics[width=\textwidth]{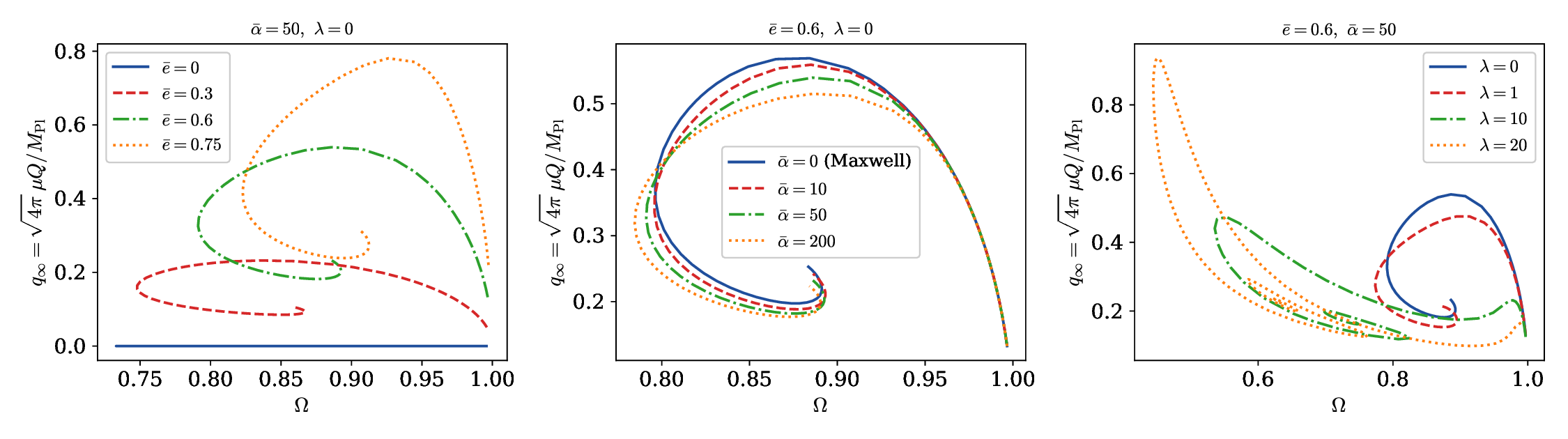}
\caption{Total mass (top) and total charge (bottom) as functions of the frequency. The first maximum of each mass curve is the turning point listed in Table~\ref{tab:max}.}
\label{fig:MOmega}
\end{figure*}

\begin{figure*}[!t]\centering
\includegraphics[width=\textwidth]{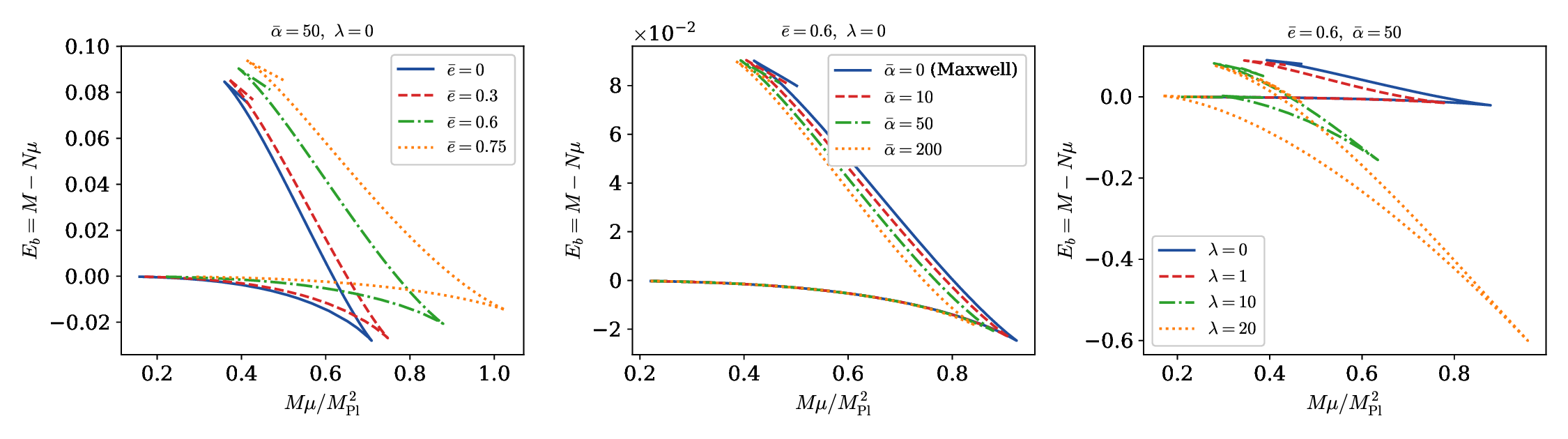}\\
\includegraphics[width=\textwidth]{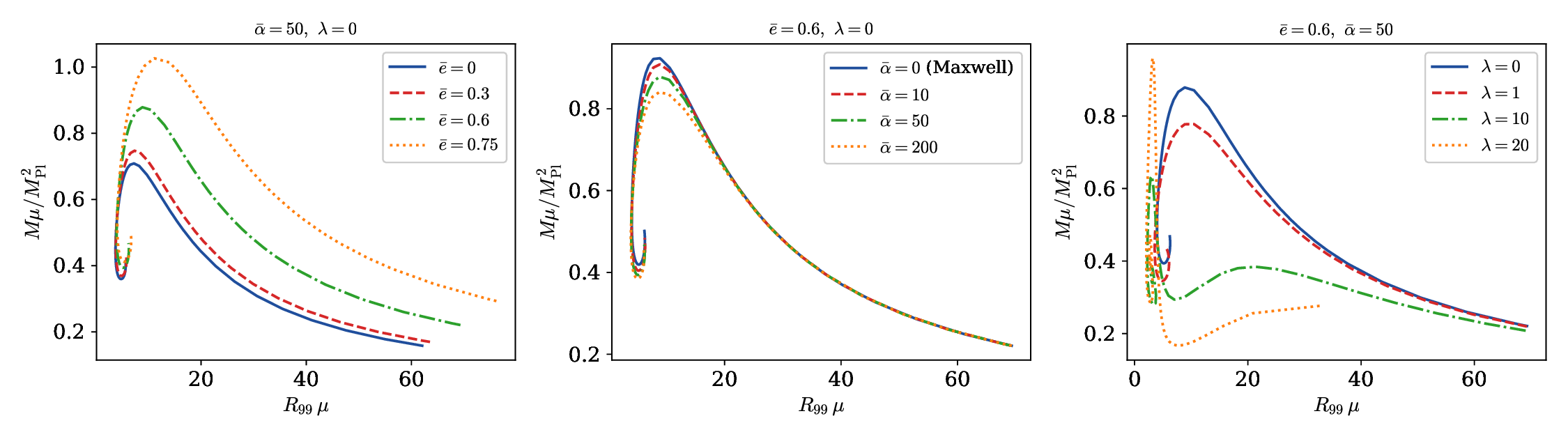}\\
\includegraphics[width=\textwidth]{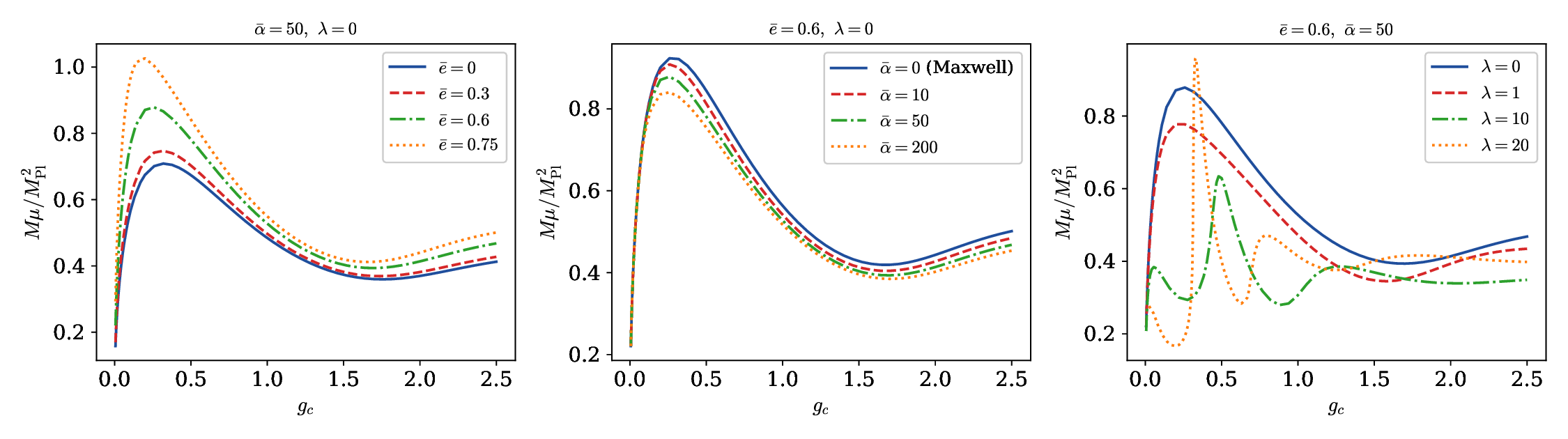}
\caption{Binding energy against mass (top), mass against radius (middle) and mass against $g_c$ (bottom).}
\label{fig:binding}
\end{figure*}

We present three groups of solutions. In the first group we fix $\bar\alpha=50$ and $\lambda=0$ and vary the charge, $\bar e=0,\,0.3,\,0.6,\,0.75$. In the second group we fix $\bar e=0.6$ and $\lambda=0$ and vary the EH coupling, $\bar\alpha=0,\,10,\,50,\,200$, where $\bar\alpha=0$ is the Maxwell case. In the third group we fix $\bar e=0.6$ and $\bar\alpha=50$ and vary the self-interaction, $\lambda=0,\,1,\,10,\,20$. Table~\ref{tab:gc05} lists the main properties of the solutions with central value $g_c=0.5$, and the radial profiles of these solutions are shown in Figs.~\ref{fig:fg}-\ref{fig:sound}. Figures~\ref{fig:MOmega} and \ref{fig:binding} show complete families obtained by varying $g_c$ from $0.006$ ($0.025$ for $\lambda=20$) to $2.5$, and Table~\ref{tab:max} lists their maximal masses. All solutions with $\bar\alpha\ge10$ have $16\bar\alpha E_{\max}^2$ between $0.3$ and $2.8$ (Table~\ref{tab:gc05}) and therefore belong to the effective quartic model.

To see what survives in the controlled regime we computed the solution with $g_c=0.5$ and $\bar e=0.6$ for small couplings, chosen so that the maximal field takes fixed values of $E_{\max}/E_c$ through Eq.~\eqref{eq:validity} (Table~\ref{tab:small}). 
For $E_{\max}\lesssim E_c$ the numerical response is approximately quadratic and linear in $\bar\alpha$,
\begin{equation}
\frac{\Delta M}{M}\simeq-1.49\times10^{-5}\left(\frac{E_{\max}}{E_c}\right)^2,
\label{eq:dMQED}
\end{equation}
with $\Delta M\simeq-6.7\times10^{-3}\bar\alpha$, $\Delta\Omega\simeq+5.3\times10^{-4}\bar\alpha$ and $\Delta q_\infty\simeq-4.6\times10^{-3}\bar\alpha$. The relative mass correction is $1.5\times10^{-9}$ at $E_{\max}/E_c=0.01$, $1.5\times10^{-7}$ at $0.1$ and $1.5\times10^{-5}$ at the critical field. All trends found below have the same sign in this regime, but their size is tiny. The quantum electrodynamic corrections to these stars are thus very small, and the large effects discussed below are properties of the quartic model at strong coupling, where $E_{\max}/E_c$ reaches $55$ for $\bar\alpha=10$.

\begin{table}[!t]
\centering
\caption{Response to the EH coupling for $g_c=0.5$, $\bar e=0.6$, $\lambda=0$. Rows with $E_{\max}\le E_c$ are controlled QED corrections; the last rows belong to the effective quartic model. $\Delta M$ and $\Delta M/M$ are measured from the Maxwell solution $M=0.8445288$.}
\label{tab:small}
\begin{tabular}{|c|c|c|c|c|c|c|c|c|c|c|c|c|c|c|}
\hline
$E_{\max}/E_c$ & $\bar\alpha$ & $16\bar\alpha E_{\max}^2$ & $\Delta M$ & $\Delta M/M$\\
\hline
0.01 & $1.87\times10^{-7}$ & $1.0\times10^{-8}$ & $-1.26\times10^{-9}$ & $-1.5\times10^{-9}$\\
\hline
0.05 & $4.68\times10^{-6}$ & $2.6\times10^{-7}$ & $-3.15\times10^{-8}$ & $-3.7\times10^{-8}$\\
\hline
0.1 & $1.87\times10^{-5}$ & $1.0\times10^{-6}$ & $-1.26\times10^{-7}$ & $-1.5\times10^{-7}$\\
\hline
0.5 & $4.68\times10^{-4}$ & $2.6\times10^{-5}$ & $-3.15\times10^{-6}$ & $-3.7\times10^{-6}$\\
\hline
1 & $1.87\times10^{-3}$ & $1.0\times10^{-4}$ & $-1.26\times10^{-5}$ & $-1.5\times10^{-5}$\\
\hline
7.3 & $10^{-1}$ & $5.4\times10^{-3}$ & $-6.6\times10^{-4}$ & $-7.8\times10^{-4}$\\
\hline
22 & 1 & $5.0\times10^{-2}$ & $-5.9\times10^{-3}$ & $-7.0\times10^{-3}$\\
\hline
55 & 10 & $3.1\times10^{-1}$ & $-3.2\times10^{-2}$ & $-3.8\times10^{-2}$\\
\hline
\end{tabular}
\end{table}

\paragraph{Spinor and metric functions.}
The spinor functions are shown in Fig.~\ref{fig:fg}. For $\lambda=0$ the charge and the EH coupling change them only slightly. The radial profiles become slightly broader as $\bar e$ grows, because the electric repulsion pushes the fermions apart. The self-interaction has a much stronger effect. For $\lambda=10$ and $20$ the attraction between the fermions concentrates the field near the center. The function $f$ then develops a tall peak at $x\approx0.7$ to $1.1$, and $R_{99}$ shrinks from $5.6$ to $2.8$ and $2.0$. This behavior is the same as in the uncharged case \cite{Dzhunushaliev2019a,Alibekova2026}. The mass function and the metric function $\sigma$ are shown in Fig.~\ref{fig:metric}. In the charged stars $m(x)$ keeps growing outside the spinor core, because the electric field stores energy there. It approaches $M-q_\infty^2/2x$ at large $x$. The electrostatic energy is $5.1\%$ of the total mass for $\bar e=0.6$ and $\bar\alpha=50$, and $6.2\%$ in the Maxwell case with the same $\bar e$.

\paragraph{Electric field and frequency.}
Figure~\ref{fig:field} shows the electric field intensity and the potential, and Fig.~\ref{fig:charge} the enclosed charge and the local frequency. The field is zero at the center, as required by regularity. It reaches its maximum near $x\approx1.8$, where most of the charge is already enclosed, and then falls as $q_\infty^2/x^4$. The middle panels show the main effect of the EH term. For the same charge coupling $\bar e=0.6$ the peak value of $E^2$ is $3.4\times10^{-3}$ in the Maxwell case and falls to $1.9\times10^{-3}$, $1.0\times10^{-3}$ and $0.5\times10^{-3}$ for $\bar\alpha=10$, $50$ and $200$. Two different effects contribute to this reduction, and we separate them. The first is the constitutive effect at fixed displacement. A given enclosed charge fixes $D$ through Eq.~\eqref{eq:gauss}, and a larger $\alpha$ lets the same $D$ be carried by a weaker $E$. The field energy at fixed $D$ also decreases, because the energy density $H(D)$ obeys $dH/dD=E<D$, so that $H(D)<D^2/2$ for $\alpha>0$. In this sense the nonlinear vacuum screens the field, although it does not change the charge itself. The second effect is the backreaction: the spinor configuration adjusts to the weaker repulsion and the enclosed charge changes. For $\bar\alpha=200$ the total field energy drops from $0.0526$ (Maxwell solution) to $0.0329$. If the charge profile $q(x)$ of the EH solution carried a Coulomb field, its energy would be $0.0426$. Hence $0.0097$ of the reduction comes from the constitutive law and $0.0100$ from the smaller enclosed charge. The two effects are of similar size, and the term screening below refers only to the first one. Locally, at the field maximum the EH energy $12\alpha E^4$ exceeds the Maxwell energy $E^2/2$ for $\bar\alpha=50$ (their ratio is $1.2$), although the total field energy is smaller than in the Maxwell star. The weaker field also lowers the potential $\phi$. The local frequency $\omega=\Omega-\phi$ is smaller than $\Omega$ in the interior, by up to $0.17$ for $\bar e=0.75$, and grows to $\Omega$ at infinity. The repulsion therefore lowers the local energy available to the fermions in the center. To remain bound, the star needs a larger asymptotic frequency. This explains why $\Omega$ grows with $\bar e$ in Table~\ref{tab:gc05}.

\paragraph{Energy density, pressures and anisotropy.}
The energy density is shown in Fig.~\ref{fig:stress}, and the main properties of the radial pressure, the tangential pressure and the anisotropy are collected in Table~\ref{tab:pressures}. For $\lambda=0$ the energy density is largest at the center and decreases monotonically. Its central value changes little with the charge ($\rho_c=0.378$ for $\bar e=0$ and $0.365$ for $\bar e=0.75$). The central pressure $p_c=p_r(0)=p_t(0)$ is positive and slightly smaller in the charged stars, and both pressures are largest at the center. In the neutral star $p_r$ stays positive and decays with the spinor. In the charged stars $p_r$ changes sign at a radius $x_0$ just outside the spinor core, because beyond it only the field remains and $p_r=-\rho_{\rm EH}$. The radius $x_0$ moves inward with the charge, from $3.24$ for $\bar e=0.3$ to $2.37$ for $\bar e=0.75$, while the most negative radial pressure, reached near $x\approx3$, grows from $-5.7\times10^{-5}$ to $-7.8\times10^{-4}$. The EH coupling has the opposite effect: from the Maxwell case to $\bar\alpha=200$ the radius $x_0$ grows from $2.38$ to $2.70$, and the field tension at its minimum drops by a factor of $2.4$. The tangential pressure is positive everywhere.

The self-interaction changes this picture. For $\lambda=10$ and $20$ the central pressures are negative ($p_c=-0.099$ and $-0.121$), and the central energy density grows to $0.58$ and $1.14$. The decomposition in Eq.~\eqref{eq:pressures} explains this. At the center the kinetic stress is $g_cf_1>0$, while the self-interaction adds $-\lambda g_c^4/2$. For $\lambda=10$ these are $+0.213$ and $-0.313$, which give the computed $p_c=-0.099$. The attractive interaction acts like a negative pressure, similar to a bag constant. For $\lambda=10$ the radial pressure stays negative at all radii, while for $\lambda=20$ it becomes positive in a thin shell $0.36<x<0.80$. In both cases the tangential pressure is large and positive, with maxima $0.111$ and $0.906$ near $x\approx1.1$ and $0.6$. These stars are therefore supported by tangential stresses rather than by a positive central pressure, and fluid-star intuition does not apply to them.

\begin{table*}[!t]
\centering
\caption{Pressures and anisotropy of the solutions with $g_c=0.5$. The central pressure is $p_c=p_r(0)=p_t(0)$. The largest tangential pressure and the largest anisotropy $\Delta=p_t-p_r$ are given with their positions in brackets. $x_0$ lists the radii where $p_r$ changes sign, and $p_{r,\min}$ is the most negative radial pressure with its position.}
\label{tab:pressures}
\begin{tabular}{|c|c|c|c|c|c|c|c|c|c|c|c|c|c|c|}
\hline
$\lambda$ & $\bar e$ & $\bar\alpha$ & $\rho_c$ & $p_c$ & $p_{t,\max}$ ($x$) & $x_0$ & $p_{r,\min}$ ($x$) & $\Delta_{\max}$ ($x$)\\
\hline
0 & 0 & $\cdot$ & 0.3779 & $+0.0426$ & 0.0426 (0.00) & none & $\cdot$ & 0.0059 (1.35)\\
\hline
0 & 0.3 & 50 & 0.3752 & $+0.0417$ & 0.0417 (0.00) & 3.24 & $-5.70\times10^{-5}$ (3.89) & 0.0064 (1.39)\\
\hline
0 & 0.6 & 50 & 0.3692 & $+0.0397$ & 0.0397 (0.00) & 2.56 & $-4.21\times10^{-4}$ (3.22) & 0.0074 (1.47)\\
\hline
0 & 0.75 & 50 & 0.3654 & $+0.0385$ & 0.0385 (0.00) & 2.37 & $-7.75\times10^{-4}$ (3.05) & 0.0082 (1.52)\\
\hline
0 & 0.6 & 0 & 0.3636 & $+0.0379$ & 0.0379 (0.00) & 2.38 & $-6.99\times10^{-4}$ (3.02) & 0.0089 (1.57)\\
\hline
0 & 0.6 & 10 & 0.3664 & $+0.0388$ & 0.0388 (0.00) & 2.45 & $-5.71\times10^{-4}$ (3.10) & 0.0081 (1.51)\\
\hline
0 & 0.6 & 200 & 0.3716 & $+0.0405$ & 0.0405 (0.00) & 2.70 & $-2.87\times10^{-4}$ (3.37) & 0.0070 (1.43)\\
\hline
1 & 0.6 & 50 & 0.3720 & $+0.0198$ & 0.0205 (0.60) & 2.19 & $-4.94\times10^{-4}$ (2.83) & 0.0101 (1.32)\\
\hline
10 & 0.6 & 50 & 0.5777 & $-0.0991$ & 0.1112 (1.14) & none & $-0.0991$ (0.00) & 0.1270 (1.10)\\
\hline
20 & 0.6 & 50 & 1.1370 & $-0.1210$ & 0.9059 (0.61) & 0.36, 0.80 & $-0.1210$ (0.00) & 0.7547 (0.62)\\
\hline
\end{tabular}
\end{table*}

The anisotropy $\Delta=p_t-p_r$ vanishes at the center and is positive elsewhere. Its maximum, listed in Table~\ref{tab:pressures}, lies near $x\approx1.4$ for $\lambda=0$ and grows with the charge, from $0.0059$ for the neutral star to $0.0082$ for $\bar e=0.75$, since the field adds $p_t^{\rm EH}-p_r^{\rm EH}=E^2+16\alpha E^4$. It decreases with $\bar\alpha$, from $0.0089$ in the Maxwell case to $0.0070$ for $\bar\alpha=200$, because the field is screened. The self-interaction raises the anisotropy strongly, to $0.127$ for $\lambda=10$ and $0.755$ for $\lambda=20$, and moves its maximum inward. Figure~\ref{fig:aniso} shows the field energy density $\rho_{\rm EH}$ and its EH part $12\alpha E^4$. The EH part is concentrated in a shell around $x\approx2$.

\paragraph{Energy conditions.}
The energy conditions are summarized in Table~\ref{tab:ec} for all ten solutions with $g_c=0.5$, including the self-interacting ones with negative central pressure. For each combination we give its smallest value inside $R_{99}$ divided by the local energy density, so that a positive entry means the condition holds with that relative margin. All entries are positive, and outside $R_{99}$ the combinations remain non-negative within numerical accuracy. The null, weak, strong and dominant energy conditions therefore hold for every configuration we computed, both for the total stress tensor and for the spinor part alone. The charge reduces the margins of the null and dominant conditions, from $0.89$ for the neutral star to $0.58$ for $\bar e=0.75$, because the field tension makes $p_r$ negative outside the core. The EH coupling increases them again, from $0.62$ in the Maxwell case to $0.72$ for $\bar\alpha=200$. For $\lambda=10$ and $20$ the conditions hold with margins near $0.5$, because the negative pressures are small compared with the energy density, $|p_c|/\rho_c\approx0.17$ and $0.11$. For the field part the conditions can be seen directly from Eq.~\eqref{eq:EHstress}, since $\rho_{\rm EH}+p_t^{\rm EH}=E^2+16\alpha E^4$, $\rho_{\rm EH}+p_r^{\rm EH}+2p_t^{\rm EH}=E^2+8\alpha E^4$ and $\rho_{\rm EH}-|p_t^{\rm EH}|=8\alpha E^4$ are all non-negative. The EH term makes the dominant energy condition hold with a margin that is absent in the Maxwell case, where $\rho_{\rm EH}=|p_t^{\rm EH}|$.

\begin{table*}[!t]
\centering
\caption{Energy conditions for the solutions with $g_c=0.5$. The columns give the smallest value, inside the radius $R_{99}$, of each combination divided by the local energy density. The null energy condition requires $\rho+p_r\ge0$ and $\rho+p_t\ge0$, the weak condition in addition $\rho\ge0$, the strong condition $\rho+p_r+2p_t\ge0$ and the dominant condition $\rho\ge|p_r|$ and $\rho\ge|p_t|$. Outside $R_{99}$ all combinations remain non-negative within the numerical accuracy of $10^{-10}$.}
\label{tab:ec}
\begin{tabular}{|c|c|c|c|c|c|c|c|c|c|c|c|c|c|c|}
\hline
$\lambda$ & $\bar e$ & $\bar\alpha$ & $\rho_c$ & $\frac{\rho+p_r}{\rho}$ & $\frac{\rho+p_t}{\rho}$ & $\frac{\rho+p_r+2p_t}{\rho}$ & $\frac{\rho-|p_r|}{\rho}$ & $\frac{\rho-|p_t|}{\rho}$\\
\hline
0 & 0 & $\cdot$ & 0.3779 & 1.003 & 1.067 & 1.136 & 0.887 & 0.886\\
\hline
0 & 0.3 & 50 & 0.3752 & 0.902 & 1.105 & 1.188 & 0.889 & 0.846\\
\hline
0 & 0.6 & 50 & 0.3692 & 0.687 & 1.108 & 1.198 & 0.687 & 0.681\\
\hline
0 & 0.75 & 50 & 0.3654 & 0.579 & 1.105 & 1.190 & 0.579 & 0.610\\
\hline
0 & 0.6 & 0 & 0.3636 & 0.658 & 1.104 & 1.243 & 0.658 & 0.620\\
\hline
0 & 0.6 & 10 & 0.3664 & 0.669 & 1.106 & 1.222 & 0.669 & 0.642\\
\hline
0 & 0.6 & 200 & 0.3716 & 0.717 & 1.109 & 1.174 & 0.717 & 0.740\\
\hline
1 & 0.6 & 50 & 0.3720 & 0.705 & 1.053 & 1.160 & 0.705 & 0.693\\
\hline
10 & 0.6 & 50 & 0.5777 & 0.458 & 0.828 & 0.485 & 0.458 & 0.536\\
\hline
20 & 0.6 & 50 & 1.1370 & 0.467 & 0.894 & 0.681 & 0.467 & 0.491\\
\hline
\end{tabular}
\end{table*}

\begin{figure*}[!t]\centering
\includegraphics[width=0.8\textwidth]{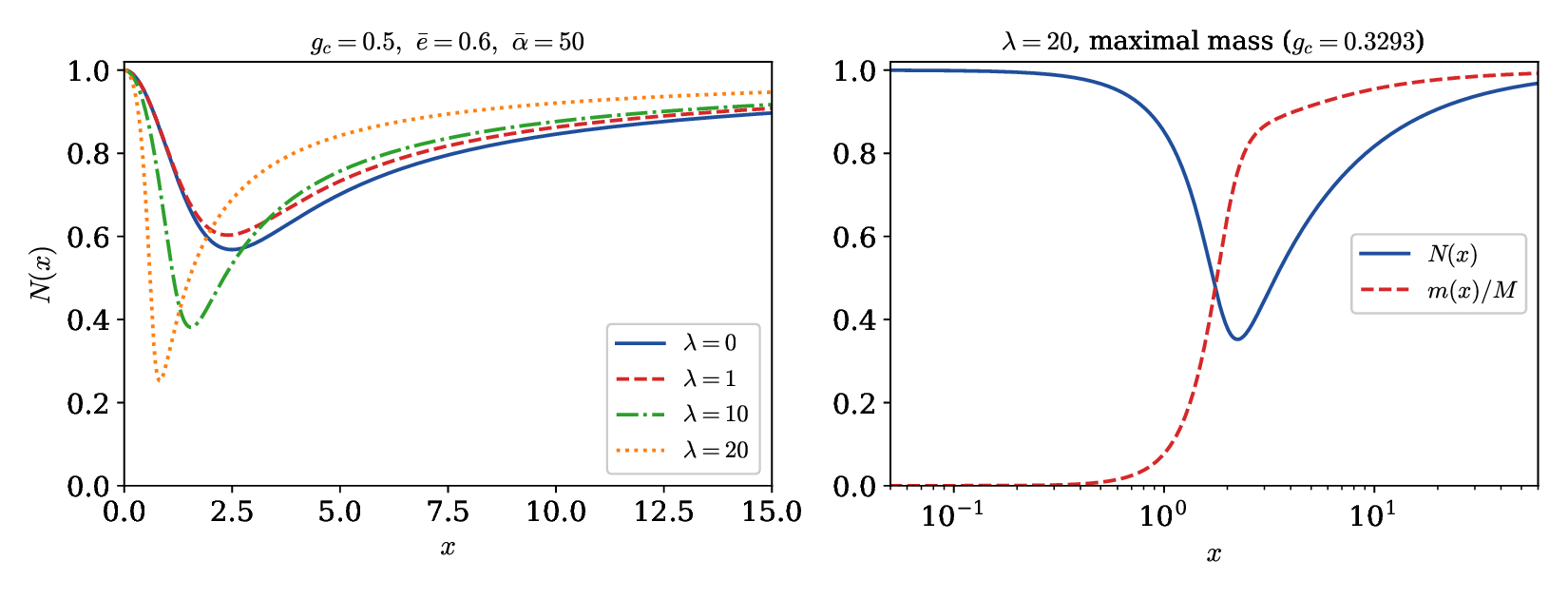}
\caption{Metric function $N(x)$ for different self-interaction at $g_c=0.5$ (left), and $N(x)$ with the mass distribution $m(x)/M$ for the $\lambda=20$ star of maximal mass (right).}
\label{fig:N}
\end{figure*}

\begin{figure}[!t]\centering
\includegraphics[width=\columnwidth]{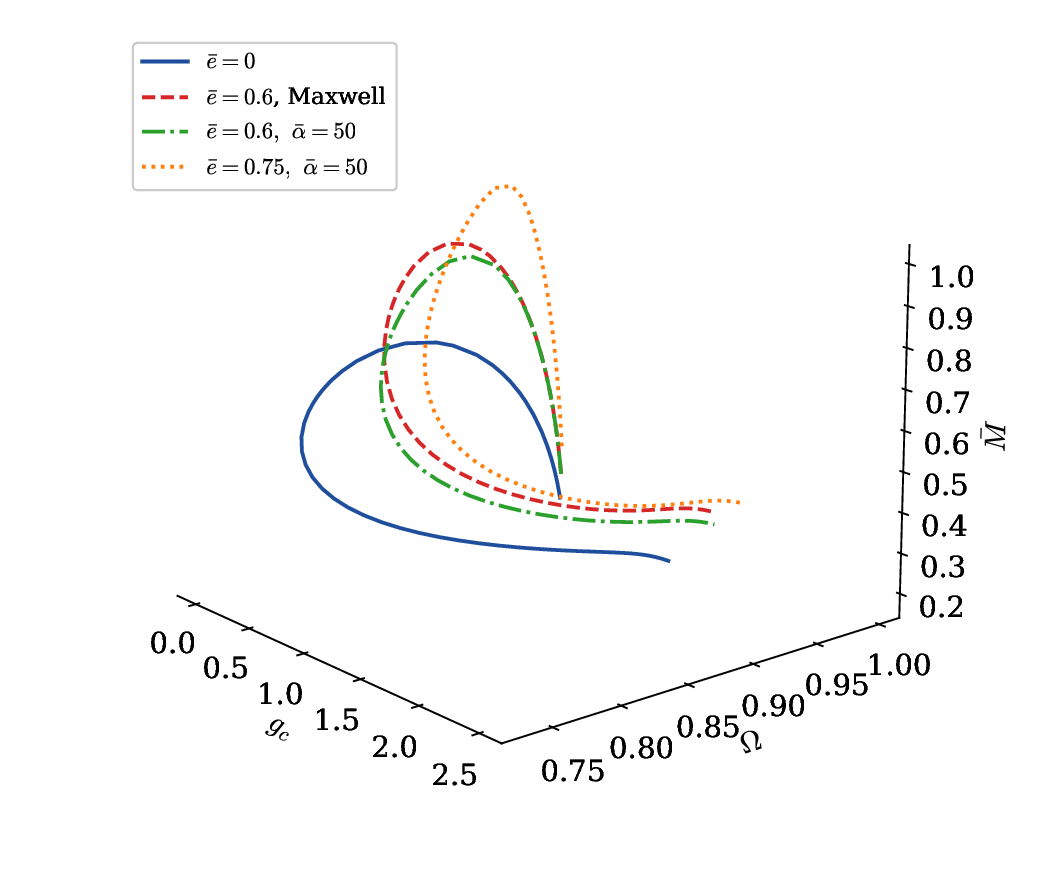}
\caption{Total mass as a function of $g_c$ and $\Omega$.}
\label{fig:3d}
\end{figure}

\paragraph{Sound speeds and adiabatic index.}
A field configuration has no equation of state, and the quantities in this paragraph are not propagation speeds of perturbations of the spinor field. We compute them only to compare with fluid models of compact stars, where they are standard diagnostics. We define $v_{r,t}^2=(dp_{r,t}/dx)/(d\rho/dx)$ along the radius. Figure~\ref{fig:sound} shows them in the core, where $p_r>0$ and $\rho$ decreases. The radial speed starts at $v_r^2\approx0.16$ at the center and decreases outward. The tangential speed stays between $0.09$ and $0.12$. Both are below one, and the charge lowers $v_t^2$ in the center. The index $\Gamma=(\rho+p_r)\,v_r^2/p_r$ stays above $4/3$ in the core and rises toward the edge, where $p_r$ goes to zero. For a fluid star $\Gamma>4/3$ is a criterion for radial stability, but it is derived under assumptions that do not hold for a coupled spinor and gauge field. Here it only shows that the effective stresses behave like those of a stiff fluid. It does not establish dynamical stability, which requires a perturbation analysis of the full system.

\begin{table*}[!t]
\centering
\caption{Maximal mass of each family and the values of $\Omega$, $g_c$, $q_\infty$, the particle number $\mathcal N$, $R_{99}$ and the minimum of the metric function $N(x)$ at the maximum. $\Omega_{\min}$ is the lowest frequency of the family.}
\label{tab:max}
\begin{tabular}{|c|c|c|c|c|c|c|c|c|c|c|c|c|c|c|}
\hline
$\lambda$ & $\bar e$ & $\bar\alpha$ & $M_{\max}$ & $\Omega$ & $g_c$ & $q_\infty$ & $\mathcal N$ & $R_{99}$ & $N_{\min}$ & $\Omega_{\min}$\\
\hline
0 & 0 & $\cdot$ & 0.70888 & 0.82638 & 0.319 & 0 & 0.7369 & 7.24 & 0.703 & 0.73298\\
\hline
0 & 0.3 & 50 & 0.74723 & 0.83559 & 0.319 & 0.23222 & 0.7741 & 7.35 & 0.695 & 0.74795\\
\hline
0 & 0.6 & 50 & 0.87860 & 0.88596 & 0.259 & 0.53952 & 0.8993 & 8.85 & 0.713 & 0.79115\\
\hline
0 & 0.75 & 50 & 1.02655 & 0.92629 & 0.199 & 0.78090 & 1.0412 & 11.09 & 0.739 & 0.82276\\
\hline
0 & 0.6 & 0 & 0.92356 & 0.88429 & 0.259 & 0.56891 & 0.9483 & 8.91 & 0.703 & 0.79629\\
\hline
0 & 0.6 & 10 & 0.90870 & 0.88499 & 0.259 & 0.55913 & 0.9320 & 8.89 & 0.706 & 0.79583\\
\hline
0 & 0.6 & 200 & 0.84047 & 0.88608 & 0.259 & 0.51509 & 0.8585 & 8.76 & 0.721 & 0.78458\\
\hline
1 & 0.6 & 50 & 0.77748 & 0.91367 & 0.199 & 0.47533 & 0.7922 & 10.50 & 0.784 & 0.77373\\
\hline
10 & 0.6 & 50 & 0.63416 & 0.55309 & 0.480 & 0.47359 & 0.7894 & 3.04 & 0.426 & 0.53566\\
\hline
20 & 0.6 & 50 & 0.95913 & 0.45105 & 0.3293 & 0.93557 & 1.5593 & 3.22 & 0.352 & 0.44440\\
\hline
\end{tabular}
\end{table*}

\paragraph{Families of solutions.}
The total mass and charge are plotted against the frequency in Fig.~\ref{fig:MOmega}. Every family starts at $\Omega\to1$ with small mass, where the stars are large and weakly bound. As $g_c$ grows, $\Omega$ decreases and the mass increases to a maximum. The curve then turns and winds into a spiral. This spiral is well known for boson, Proca and Dirac stars \cite{Liebling2023,Herdeiro2017} and was also found for charged Dirac stars in the Maxwell theory \cite{Hernandez2026}. The charge moves the whole spiral to larger $\Omega$ and larger $M$. As Table~\ref{tab:max} shows, the maximal mass grows from $0.709$ at $\bar e=0$ to $0.747$, $0.879$ and $1.027$ for $\bar e=0.3$, $0.6$ and $0.75$. The frequency at the maximum grows from $0.826$ to $0.926$, and the radius at the maximum grows from $7.2$ to $11.1$. The repulsion lets the star hold more particles before gravity wins, but it also makes the star larger.

The EH term acts in the opposite direction. For $\bar e=0.6$ the maximal mass decreases from $0.924$ in the Maxwell case to $0.909$, $0.879$ and $0.840$ for $\bar\alpha=10$, $50$ and $200$. This is a reduction of $9\%$, and the charge at the maximum falls from $0.569$ to $0.515$. The frequency at the maximum hardly changes.

Since the total charge is $q_\infty=\bar e\mathcal N$, a comparison at fixed $\bar e$ mixes the nonlinear response with a change of the particle number. We therefore also compare the stars at fixed $\mathcal N$, which is the same as fixed total charge (Table~\ref{tab:fixedN}). At fixed $\mathcal N$ the EH term changes the mass very little: for $\mathcal N=0.85$ the mass decreases by $0.001$ ($0.12\%$) from the Maxwell value at $\bar\alpha=200$. The stars become more compact, with $R_{99}$ falling from $13.3$ to $10.6$, their frequency is lower and they are slightly more bound. The large reduction of $M_{\max}$ has a different origin. The weaker field repels the fermions less, so the turning point of the family is reached at a smaller particle number: $\mathcal N_{\max}=0.948$ in the Maxwell case and $0.932$, $0.899$ and $0.859$ for $\bar\alpha=10$, $50$ and $200$. In other words, the nonlinear term lowers the largest charge and particle number that a stable star can carry, rather than the mass of a given star.

\begin{table}[!t]
\centering
\caption{Stars with the same particle number $\mathcal N$ (and therefore the same total charge) on the branch before the maximal mass, $\bar e=0.6$, $\lambda=0$.}
\label{tab:fixedN}
\begin{tabular}{|c|c|c|c|c|c|c|c|c|c|c|c|c|c|c|}
\hline
$\mathcal N$ & $\bar\alpha$ & $M$ & $E_b$ & $\Omega$ & $R_{99}$\\
\hline
0.50 & 0 & 0.49709 & $-0.00291$ & 0.9819 & 28.74\\
\hline
0.50 & 200 & 0.49709 & $-0.00291$ & 0.9818 & 28.69\\
\hline
0.70 & 0 & 0.69156 & $-0.00844$ & 0.9606 & 18.59\\
\hline
0.70 & 200 & 0.69148 & $-0.00852$ & 0.9594 & 18.16\\
\hline
0.85 & 0 & 0.83369 & $-0.01631$ & 0.9320 & 13.26\\
\hline
0.85 & 10 & 0.83363 & $-0.01637$ & 0.9309 & 13.12\\
\hline
0.85 & 50 & 0.83344 & $-0.01656$ & 0.9268 & 12.56\\
\hline
0.85 & 200 & 0.83272 & $-0.01728$ & 0.9100 & 10.61\\
\hline
\end{tabular}
\end{table} The effect is limited to the part of the family with strong fields. Near $\Omega\to1$ all curves with the same $\bar e$ coincide, because there $16\alpha E^2\ll1$ and the EH term is negligible. The self-interaction lowers the maximal mass to $0.777$ for $\lambda=1$. For $\lambda=10$ the family reaches much lower frequencies ($\Omega_{\min}=0.536$) and develops additional loops, as in the uncharged case \cite{Dzhunushaliev2019a,Alibekova2026}. For $\lambda=20$ the mass rises very steeply in a narrow interval $0.30\lesssim g_c\lesssim0.33$ and reaches $M_{\max}=0.95913$ at $g_c=0.3293$ and $\Omega=0.45105$. Because of this steep variation we resolved the maximum with a fine arclength step and checked its convergence separately (Appendix~\ref{app:numerics}).

The maximal mass depends on $\lambda$ in a nonmonotonic way: it falls from $0.879$ ($\lambda=0$) to $0.777$ ($\lambda=1$) and $0.634$ ($\lambda=10$), and then rises to $0.959$ ($\lambda=20$). The same behavior is found for uncharged Dirac stars in $R^2$ gravity \cite{Alibekova2026}, where the maximal mass for $\lambda=20$ also exceeds the one for $\lambda=10$. Writing $M=\mathcal N+E_b$ makes the reason clear. For $\lambda\le1$ the stars at the maximum are extended and weakly bound, $E_b/\mathcal N\approx-0.02$, and the attraction mainly makes them more compact, so that the turning point is reached with fewer particles. For $\lambda=10$ and $20$ the attraction dominates the binding, with $E_b/\mathcal N=-0.20$ and $-0.38$ and frequencies $\Omega\approx0.55$ and $0.45$. These are compact, self-bound objects, and the turning point moves to a much larger particle number, $\mathcal N=0.79$ and $1.56$. For $\lambda=20$ the gain in particle number outweighs the stronger binding, and the maximal mass increases again.

The $\lambda=20$ star at the maximum is also very compact, with $2M/R_{99}\approx0.60$. This ratio does not indicate a horizon, because $R_{99}$ is not a surface and a large part of the mass lies in the electric field outside it ($R_M=45.6$). Figure~\ref{fig:N} shows the metric function $N(x)=1-2m/x$ and the mass distribution of this star. The minimum $N_{\min}=0.352$ is reached at $x=2.23$, so no horizon forms. Along all computed families $N(x)$ stays above $0.21$, and Table~\ref{tab:max} lists $N_{\min}$ at each maximal mass. At $g_c=0.5$, $N_{\min}$ lies between $0.55$ and $0.60$ for $\lambda\le1$ and equals $0.38$ and $0.25$ for $\lambda=10$ and $20$.

The binding energy, the mass-radius relation and the mass as a function of $g_c$ are shown in Fig.~\ref{fig:binding}. The binding energy has the cusp structure that is typical of solitons. Along the branch from $\Omega\to1$ to the maximal mass the stars are bound ($E_b<0$). Beyond the cusp at $M_{\max}$ the binding energy grows and becomes positive, and such stars could release energy by dispersing. We distinguish three statements. The bound branch is the part of each family with $E_b<0$. The first maximum of $M(\Omega)$ is a turning point, where for boson stars a radial mode is known to change stability \cite{Liebling2023}. By analogy, the branch between $\Omega\to1$ and the first maximum is the candidate for stable charged Dirac stars. We have not performed a perturbation analysis, so this remains a heuristic indicator and no branch is shown here to be dynamically stable. The solution with $\bar e=0.75$ and $g_c=0.5$ in Table~\ref{tab:gc05} has $E_b>0$ because it lies beyond the turning point of its family, which occurs at $g_c=0.199$. The charge makes the bound branch shallower. The lowest binding energy is $-0.028$ for $\bar e=0$ and $-0.015$ for $\bar e=0.75$. Bound stars still exist for $\bar e=0.75$, in agreement with the result of Ref.~\cite{Hernandez2026} that bound charged Dirac stars exist below $\bar e=1$. The EH term moves the cusp to smaller mass but leaves the bound branch close to the Maxwell one (lowest $E_b=-0.021$ for $\bar\alpha=50$ against $-0.025$ for Maxwell). In the mass-radius plane the stars become larger and heavier with the charge, and the EH term reduces both. A Dirac star has no sharp surface, and in the charged case the field outside the spinor core carries part of the mass. Table~\ref{tab:gc05} therefore lists two radii. The radius $R_M$ that contains $99\%$ of the mass is close to $R_{99}$ for the neutral star ($R_M/R_{99}=0.96$) but much larger for the charged ones, because the field energy falls only as $q_\infty^2/2x$. The ratio grows with the charge, $R_M/R_{99}=1.09$, $2.49$ and $3.95$ for $\bar e=0.3$, $0.6$ and $0.75$ at $\bar\alpha=50$, and decreases with the EH coupling, from $2.71$ in the Maxwell case to $2.42$ for $\bar\alpha=200$. For the compact self-interacting stars it reaches $6.2$ ($\lambda=10$), since their spinor core is small while the field extends far outside it. The ratio is therefore a simple measure of how much of the gravitating mass sits in the electric field rather than in the matter. The ratio $2M/R_{99}$, between $0.25$ and $0.29$ for $\lambda=0$ and $0.45$ for $\lambda=10$, is a measure of how concentrated the spinor is and should not be read as the surface compactness of a fluid star. A three-dimensional view of the families is given in Appendix~\ref{app:numerics}.

\section{Conclusions}\label{sec:conclusions}

We have constructed Dirac stars made of a charged spinor field whose electric field obeys the EH Lagrangian. The model combines three ingredients that were studied separately before: the two fermion Dirac stars of general relativity, the gauge coupling of charged Dirac stars and the nonlinear vacuum response of QED. We derived the field equations and showed that the EH term enters through a cubic constitutive relation between the field and the enclosed charge, which can be solved in closed form. The exterior of the stars is the EH-corrected Reissner-Nordstr\"om geometry. We solved the coupled field equations numerically, and the code reproduces the benchmark results for uncharged Dirac stars to numerical accuracy.

Our main findings are the following. The electric repulsion lowers the local frequency of the fermions in the center. As a result the stars need a larger frequency, they become larger and heavier, and their binding weakens. The maximal mass increases from $0.709$ to $1.027$ as $\bar e$ grows from $0$ to $0.75$.

The central result concerns the effect of the quartic term on stars with a fixed number of fermions, and therefore a fixed total charge. Such stars differ in mass by less than $0.2\%$ from their Maxwell counterparts, but they are more compact and more strongly bound, because the nonlinear field law weakens the electric repulsion. As a consequence the turning point of the family moves to a smaller particle number, and this shift, not a change in the mass of a given star, is what lowers the maximal mass and charge, by $9\%$ at $\bar e=0.6$ and $\bar\alpha=200$. About half of the drop of the field energy comes from the nonlinear constitutive law and half from the change of the enclosed charge. These large effects belong to the quartic model at strong coupling. Within the QED weak-field regime the correction is $\Delta M/M\simeq-1.5\times10^{-5}(E_{\max}/E_c)^2$, too small to matter for the structure of the stars.

All solutions satisfy the null, weak, strong and dominant energy conditions and have a positive anisotropy that grows with the charge. The attractive self-interaction makes the stars much more compact and produces negative central pressures. These stars are supported by tangential stresses rather than by a positive central pressure. Their maximal mass depends on $\lambda$ in a nonmonotonic way, because strongly self-bound stars reach the turning point with many more particles. No horizon forms along any computed family, and the metric function stays above $0.21$. The binding energy and the turning point of the mass curve identify the branch before the maximal mass as the candidate for stability, but this has not been proven here.

Several extensions are natural. A linear analysis of radial perturbations of the metric, the spinor and the gauge field would settle the stability of the families. The full EH effective action would extend the results beyond the weak-field regime. Rotating solutions and the inclusion of a magnetic dipole would connect the model to magnetized compact objects.

\section*{Acknowledgements}
SC acknowledges the Istituto Nazionale di Fisica Nucleare (INFN) Sez. di Napoli,  Iniziative Specifiche QGSKY and MoonLight-2  and the Istituto Nazionale di Alta Matematica (INdAM), gruppo GNFM, for the support. This paper is based upon work from COST Action CA21136 - Addressing observational tensions in cosmology with systematics and fundamental physics (CosmoVerse), supported by COST (European Cooperation in Science and Technology).

\appendix

\section{Numerical method and convergence}\label{app:numerics}

The shooting integration uses an explicit Runge-Kutta method of order five with relative and absolute tolerances $10^{-10}$ and $10^{-12}$, starting at $x_0=10^{-4}$ from the expansion of Appendix~\ref{app:center}. The frequency is bisected 48 times. A trial frequency is classified as too large when $g$ crosses zero, and as too small when $N$ approaches zero or the spinor grows. The fundamental solution is the nodeless one, in which $g$ has no zero. The collocation solver \cite{Virtanen2020} works on $[10^{-3},x_b]$ with relative tolerance $10^{-6}$ and an adaptive mesh of up to $1.5\times10^5$ points (typically $700$). The boundary conditions are $f=f_1x_0$, $g=g_c$, $m=m_3x_0^3$, $\sigma=1$, $\phi=0$ and $q=q_3x_0^3$ at $x_0$, while $\Omega$ is an unknown parameter. At the outer boundary we impose the asymptotic decay of the spinor instead of setting a component to zero. With $N$, $\sigma$ and $\phi$ frozen at $x_b$, Eqs.~\eqref{eq:f} and \eqref{eq:g} give $f,g\propto e^{-kx}$ with
\begin{equation}
k^2=\frac1{N}\left(1-\frac{\omega^2}{N\sigma^2}\right),\qquad kf=\left(\frac{1}{\sqrt N}-\frac{\omega}{N\sigma}\right)g ,
\label{eq:robin}
\end{equation}
and the second relation is used as a Robin condition at $x_b$. Replacing it by the simpler condition $g(x_b)=0$ changes $M$, $\Omega$ and $q_\infty$ by less than $10^{-8}$ for all tested solutions, including the extended star with $g_c=0.05$ and the self-interacting star with $\lambda=10$, because the spinor has decayed by many orders of magnitude at $x_b$. The outer radius is $x_b=\max(80,\,35/\sqrt{1-\Omega^2})$, up to $x_b\approx460$ at the smallest $g_c$. Families are traced with an adaptive step in $g_c$ that is halved when the solver fails or when $M$ or $\Omega$ jumps by more than $0.05$. For $\lambda=20$, where the mass changes very rapidly for $0.30\lesssim g_c\lesssim0.33$, we use pseudo-arclength continuation in $(g_c,\Omega)$ with steps of at most $1.5\times10^{-3}$ near the maximum. At the maximum ($g_c=0.32925$) the values $M=0.9591267$, $\Omega=0.4510532$ and $q_\infty=0.9355704$ are identical for tolerances $10^{-6}$ and $10^{-8}$, for $x_b=80$ and $160$, and for the Robin and Dirichlet outer conditions, while $N_{\min}=0.3522$ in all cases. Table~\ref{tab:conv} shows that the results do not depend on the outer radius, also for an extended star close to $\Omega=1$. They also do not depend on the collocation resolution. For $g_c=0.5$, tightening the tolerance from $10^{-6}$ to $10^{-8}$ (712 and 1059 mesh points) changes $M$ by $4\times10^{-9}$, and doubling the mesh of the $10^{-8}$ solution leaves $M=0.781290105$, $\Omega=0.822888912$ and $q_\infty=0.470447190$ unchanged in all digits.

\begin{table}[!h]
\centering
\caption{Dependence on the outer radius $x_b$ ($\bar e=0.6$, $\bar\alpha=50$, $\lambda=0$).}
\label{tab:conv}
\begin{tabular}{|c|c|c|c|c|c|c|c|c|c|c|c|c|c|c|}
\hline
$g_c$ & $x_b$ & $M$ & $\Omega$ & $q_\infty$\\
\hline
0.5 & 40 & 0.7812901 & 0.8228890 & 0.4704472\\
\hline
0.5 & 80 & 0.7812901 & 0.8228889 & 0.4704472\\
\hline
0.5 & 160 & 0.7812901 & 0.8228889 & 0.4704472\\
\hline
0.05 & 150 & 0.5888198 & 0.9733743 & 0.3562888\\
\hline
0.05 & 300 & 0.5888197 & 0.9733743 & 0.3562887\\
\hline
0.05 & 600 & 0.5888194 & 0.9733743 & 0.3562885\\
\hline
\end{tabular}
\end{table}

\section{Expansion near the center}\label{app:center}

We insert $f=f_1x+O(x^3)$, $g=g_c+g_2x^2+O(x^4)$, $m=m_3x^3+O(x^5)$, $\sigma=\sigma_c+O(x^2)$ and $q=q_3x^3+O(x^5)$ into Eqs.~\eqref{eq:f}-\eqref{eq:em} and write $w_c=\omega_c/\sigma_c$. Near the center $N\simeq1-2m_3x^2$, so that $N'/4N\simeq-m_3x$ and $(1-1/\sqrt N)/x\simeq-m_3x$, while $P\simeq-g_c^2$. At order $x^0$ Eq.~\eqref{eq:f} gives $3f_1+g_c(1-w_c-\lambda g_c^2)=0$, so
\begin{equation}
f_1=\frac{g_c}{3}\left(w_c-1+\lambda g_c^2\right).
\end{equation}
The central energy density follows from Eq.~\eqref{eq:m} with $n=g_c^2$, $P^2=g_c^4$ and $E(0)=0$, so that
\begin{equation}
m_3=\frac{\rho_c}{3}=\frac13\left(w_cg_c^2+\frac{\lambda}{2}g_c^4\right).
\end{equation}
Equation~\eqref{eq:sigma} gives $\sigma'/\sigma\simeq x(w_cg_c^2+g_cf_1)$, and Eq.~\eqref{eq:qphi} gives $q_3=eg_c^2/3$ and $\phi\simeq\phi_c-e\sigma_cg_c^2x^2/6$. At order $x$ Eq.~\eqref{eq:g} then yields
\begin{eqnarray}
g_2&=&-\frac12\Big[g_c\Big(-2m_3+\frac{w_cg_c^2+g_cf_1}{2}\Big)\nonumber\\
&&\qquad+f_1\left(1+w_c-\lambda g_c^2\right)\Big].
\end{eqnarray}
The electric field enters the metric only at order $x^5$ in $m$. Fitting the numerical solutions near the center reproduces these coefficients. For $\bar e=0.6$, $\bar\alpha=50$ we obtain $f_1=0.079478$ and $g_2=-0.088012$ from both the formulas and the fit, and for $\lambda=10$ we obtain $f_1=0.426783$ and $g_2=0.130203$ against the fitted $0.426782$ and $0.130202$.

\end{document}